\documentclass[11pt]{article}
\usepackage{geometry}                 
\usepackage{setspace}
\usepackage[parfill]{parskip}     
\usepackage{newtxtext}

\usepackage[utf8]{inputenc}  
\usepackage[T1]{fontenc}     
\usepackage{hyperref}        
\usepackage{url}             
\usepackage{booktabs}      
\usepackage{amsfonts}       
\usepackage{nicefrac}        
\usepackage{microtype}       
\usepackage{xcolor}         

\usepackage{amsmath}
\usepackage{amsthm}
\usepackage{amssymb}
\usepackage{mathtools} 
\usepackage{bm}
\usepackage{braket}
\usepackage{lineno}
\usepackage{pifont}
\usepackage{multirow}
\usepackage{subfigure}
 
\usepackage{diagbox}
\usepackage{graphicx}
\usepackage[normalem]{ulem}

\usepackage{adjustbox}

\DeclareMathOperator{\Tr}{Tr}
\newcommand{\XGate}{\mathop{\text{X}}}
\newcommand{\YGate}{\mathop{\text{Y}}}
\newcommand{\ZGate}{\mathop{\text{Z}}}
\newcommand{\CZ}{\mathop{\text{CZ}}}

\newcommand{\RX}{\mathop{\text{RX}}}
\newcommand{\RY}{\mathop{\text{RY}}}
\newcommand{\RZ}{\mathop{\text{RZ}}}
\newcommand{\CNOT}{\mathop{\text{CNOT}}}
\newcommand{\Hada}{\mathop{\text{H}}}

\DeclareMathOperator{\Real}{Re}

\title{\LARGE{The dilemma of quantum neural networks}}

\author{Yang Qian~  \thanks{School of Computer Science, The University of Sydney} \thanks{JD Explore Academy}   \and Xinbiao Wang~\thanks{Institute of Artificial Intelligence, School of Computer Science, Wuhan University} \footnotemark[2] \and Yuxuan Du \thanks{Corresponding author, duyuxuan123@gmail.com} \footnotemark[2] \and Xingyao Wu \footnotemark[2]    \and Dacheng Tao~\footnotemark[2]  }

\date{}

\begin{document}

\maketitle

\begin{abstract}
The core of quantum machine learning is to devise quantum models with good trainability and low generalization error bound than their classical counterparts to ensure better reliability and interpretability. Recent studies confirmed that quantum neural networks (QNNs) have the ability to achieve this goal on specific datasets. With this regard, it is of great importance to understand whether these advantages are still preserved on real-world tasks. Through systematic numerical experiments, we empirically observe that current QNNs fail to provide any benefit over classical learning models. Concretely, our results deliver two key messages. First, QNNs suffer from the severely limited effective model capacity, which incurs poor generalization on real-world datasets. Second, the trainability of QNNs is insensitive to regularization techniques, which sharply contrasts with the classical scenario. These empirical results force us to rethink the role of current QNNs and to design novel protocols for solving real-world problems with quantum advantages.
 
\end{abstract} 

\maketitle

\section{Introduction}

The theme of deep learning is efficiently optimizing a good neural network architecture with low \textit{generalization} error such that it can well extrapolate the underlying rule from the training data to new unseen data \cite{goodfellow2016deep,mohri2012foundations,vapnik2013nature}. During the past decades, deep neural networks (DNNs) with diverse architectures have been carefully designed to accomplish different tasks with both low train and test error. Moreover, these DNNs have achieved state-of-the-art performance compared with conventional machine learning models such as support vector machines \cite{mohri2012foundations}. Concrete examples include the exploitation of convolutional neural networks to tackle computer vision tasks \cite{huang2017densely,krizhevsky2012imagenet} and the employment of recurrent neural networks to solve natural language processing tasks \cite{hochreiter1997long,schuster1997bidirectional}. Alongside the huge empirical success of deep learning, numerous studies have been dedicated to investigating the excellent trainability and generalization ability of DNNs   \cite{Neyshabur2017nips,sun2019optimization,zhang2017rethink}, since a good understanding of these two properties does not only contribute to make DNNs more interpretable, but it might also lead to more reliable model architecture design. 
 						
A milestone in the regime of quantum computing is Google's experimental demonstration that modern quantum machines can solve certain computational tasks faster than classical computers \cite{arute2019quantum,preskill2018quantum}. Such a superior power fuels a growing interest of designing quantum machine learning (QML) models, which can be effectively executed on both noisy intermediate-scale quantum (NISQ) and fault-tolerant quantum machines with provable advantages \cite{biamonte2017quantum,cerezo2020variational2,harrow2017quantum,schuld2014quest}. Following this routine, the quantum neural networks (QNNs), as the quantum extension of DNNs, has been extensively investigated \cite{beer2020training,cong2019quantum,du2021grover,farhi2018classification,havlivcek2019supervised,mitarai2018quantum,schuld2019quantum}. Celebrated by their flexible structures, experimental studies have implemented QNNs on different NISQ platforms to accomplish various learning tasks such as data classification \cite{havlivcek2019supervised,kusumoto2019experimental}, image generation \cite{huang2020experimental,rudolph2020generation,zhu2019training}, and electronic-structure problems in material science and condensed matter physics \cite{hempel2018quantum,kandala2017hardware,google2020hartree,peruzzo2014variational}.

\begin{figure*}
	\centering \includegraphics[width=0.9\textwidth]{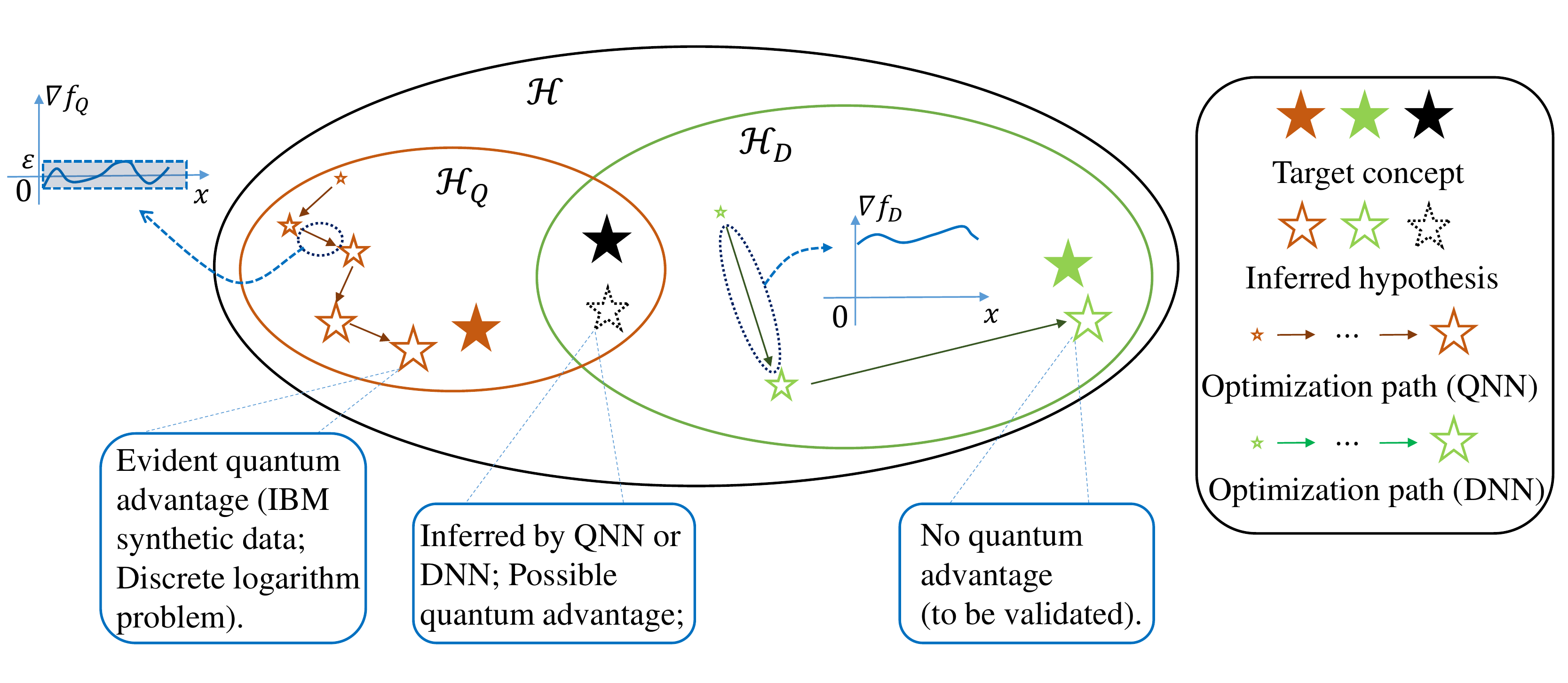}
	\caption{\small{\textbf{An overview of the classical and quantum learning models}. Generalization ability: $\mathcal{H}$ is the whole hypothesis space. $\mathcal{H}_D$ and $\mathcal{H}_Q$ refer to the hypothesis space represented by QNN and DNN respectively. When the target concept is covered by $\mathcal{H}_Q \backslash \mathcal{H}_D$ ($\mathcal{H}_D \backslash \mathcal{H}_Q$), as highlighted by the red (green) hollow star, QNNs can definitely   (fail to) guarantee computational advantages over DNNs. When the target concept lies in $\mathcal{H}_D \cap \mathcal{H}_Q$ (highlighted by the black color), it is unknown whether QNNs may possess any advantage over DNNs. Trainability: The function graph corresponding to the arrow of the optimization path of QNN indicates the possible barren plateau which is characterized by the vanished gradients.}}
	\label{fig:schem}
\end{figure*}

Driven by the promising empirical achievements of QML and the significance of understanding the power of QNNs, initial studies have been conducted to explore the trainability and the generalization ability of QNNs \cite{abbas2020power,banchi2021generalization, bu2021statistical,du2020learnability,du2021efficient, huang2020power,huang2021information} by leveraging varied model complexity measures developed in statistical learning theory \cite{vapnik1992principles}  and advanced tools in  optimization theory \cite{boyd2004convex}. Notably, the obtained results transmitted both positive and negative signals, as indicated in Figure \ref{fig:schem}. To be more concrete, theoretical evidence validated that QNNs can outperform DNNs for  specific learning tasks, i.e., quantum synthetic data classification  \cite{havlivcek2019supervised} and discrete logarithm problem \cite{huang2020power}. However, Ref. \cite{mcclean2018barren} revealed the barren plateaus' issue of QNNs, which challenges the applicability of QNNs on large-scale problems. Considering that an ambitious aim of QNNs is providing computational advantages over DNNs on real-world tasks, it is important to answer: `\textit{Are current QNNs sufficient to solve certain real-world problems with potential advantages?}' If the response is negative, it is necessary to figure out `\textit{how is the gap between QNNs and DNNs?}'   

\medskip
\noindent\textbf{Problem setup.}
 We inherit the tradition in DNNs to understand the trainability and generalization of QNNs \cite{allen2019learning}. Particularly, the explicit form of the measure of the \textit{generalization error bound}  is
\begin{eqnarray}\label{eqn:gene-error}
\hat{\mathcal{R}}_S(\hat{\bm{\theta}}) - \mathcal{R}(\hat{\bm{\theta}}): = &   \frac{1}{n} \sum_{i=1}^n C \left(h(\hat{\bm{\theta}}, \bm{x}^{(i)}), \bm{y}^{(i)} \right) \nonumber    - \mathbb{E}_{\bm{x}, \bm{y}}\left(C(h(\hat{\bm{\theta}},\bm{x}), \bm{y}) \right), 
\end{eqnarray}
where $S=\{ (\bm{x}^{(i)}, \bm{y}^{(i)})\}_{i=1}^n$ denotes the given training  dataset sampled from the domain $\mathcal{X}\times \mathcal{Y}$, $h(\hat{\bm{\theta}},\cdot)\in \mathcal{H}$ refers to the hypothesis inferred by QNN with $\mathcal{H}$ being the hypothesis space and $\hat{\bm{\theta}}$ being the trained parameters, $C: \mathcal H \times \mathcal (\mathcal X \times \mathcal Y) \to \mathbb R^+$ is the designated loss function, and $\hat{\mathcal{R}}_S(\hat{\bm{\theta}})$ (or $\mathcal{R}(\hat{\bm{\theta}})$) represents the empirical (or expected) risk \cite{kawaguchi2017generalization}. The generalization error bound in Eqn.~(\ref{eqn:gene-error}) concerns when and how minimizing $\hat{\mathcal{R}}_S(\hat{\bm{\theta}})$ is a sensible approach to minimizing $\mathcal{R}(\hat{\bm{\theta}})$. A low error bound suggests that the unearthed rule $h(\hat{\bm{\theta}})$ from the dataset $S$ can well generalize to the unseen data sampled from the same domain. Note that since the probability distribution behind data domain is generally inaccessible, the term $\mathcal R(\hat{\bm{\theta}})$ is intractable. A generic strategy is employing  the test dataset $\tilde{S}\sim \mathcal{X}\times \mathcal{Y}$ to estimate this term, i.e., $\mathcal{R}(\hat{\bm{\theta}}) \approx \frac{1}{\tilde{n}} \sum_{i=1}^{\tilde{n}} \ell (h(\hat{\bm{\theta}}, \tilde{\bm{x}}^{(i)}), \tilde{\bm{y}}^{(i)})$ with $(\tilde{\bm{x}}^{(i)}), \tilde{\bm{y}}^{(i)}) \in \tilde{S}$.

The \textit{trainability} concerns the convergence rate of the trained parameters of QNN towards the optimal parameters.  The mathematical form of the optimal parameters $\bm{\theta}^*$ satisfies 
\begin{equation}\label{eqn:QNN-overview}
	\bm{\theta}^* = \arg\min_{\bm{\theta}} \hat{\mathcal{R}}_S({\bm{\theta}}) = \arg\min_{\bm{\theta}} \frac{1}{n} \sum_{i=1}^n C \left(h({\bm{\theta}}, \bm{x}^{(i)}), \bm{y}^{(i)} \right). 
\end{equation}   
Intuitively, the inferred hypothesis (or equivalently, the trained parameters) is expected to achieve the minimized empirical risk $\hat{\mathcal{R}}_S(\bm{\theta})$. Considering that the loss landscape of QNNs is generally non-convex and non-concave, which implies the computational hardness of seeking $\bm{\theta}^*$, an alternative way to examine the trainability of QNN is analyzing its convergence rate, i.e.,
\begin{equation}\label{eqn:trainability}
	\mathcal{J}( \bm{\theta }) = \mathbb{E}[\|\nabla_{\bm{\theta}} \hat{\mathcal{R}}_S( \bm{\theta }) \|],
\end{equation}         
 where the expectation is taken over the randomness from the sample error and gate noise \cite{du2020learnability}. In other words, the metric $\mathcal{J}( \bm{\theta })$ evaluates how far the trainable parameters of QNN are away from the stationary point $\|\nabla_{\bm{\theta}} \hat{\mathcal{R}}_S( \bm{\theta }) \| = 0$.

Following the above explanations, understanding the learnability of QNNs amounts to exploring whether QNNs possess better generalization ability than DNNs on certain real-world datasets in terms of  Eqn.~(\ref{eqn:gene-error}) under both the noiseless and NISQ scenarios. Furthermore, it is crucial to understand whether the trainability of QNNs can be enhanced by regularization techniques measured by Eqn.~(\ref{eqn:trainability}).

\textbf{Contributions.} Through systematic numerical simulations, we empirically exhibit the dilemma of QNNs such that it is hard to directly use current QNNs to gain quantum advantages on real-world datasets. Meanwhile, current QNNs suffers from the poor trainability. The main contributions of our study are as follows. 

\begin{enumerate}
	\item We compare the performance of QNNs and DNNs on \textit{larger-scale datasets} than those used in previous literature to quantify the trainability and generalization ability of QNNs under both the noiseless and NISQ scenarios. As exhibited in Figure~\ref{fig:schem}, we observe the poor model capacity of QNNs by conducting randomization experiments proposed by \cite{zhang2017rethink}. Since the effective model capacity determines the model's generalization error bounds, our results suggest that the generalization error bounds of QNNs achieved by statistical learning theory are generally tight  \cite{abbas2020power,banchi2021generalization, bu2021statistical,du2021efficient,huang2020power,huang2021information}. In addition, QNNs do not gain obvious trainability enhanced assisted by regularization techniques, which sharply differs from DNNs. These observations partially explain why current QNNs fail to surpass DNNs on real-world tasks.

	\item We indicate the negative role of noise with respect to the generalization ability and trainability of QNNs. Specifically, quantum noise degenerates the model capacity  and exacerbates the difficulty of optimization. To this end, we discuss possible solutions such as advanced error mitigation techniques to enhance the capability of QNNs on real-world datasets.  
	 
	\item We build a benchmark to evaluate the performance of QNNs and DNNs on both quantum synthetic data and classical data, supporting a variety of predefined models of QNNs and providing flexible interface for researchers to define customizable architectures. The released benchmark will facilitate the standardization of assessment of various QNNs in QML community and provide a comparable reference in the design of QNNs. The related code will be released to the Github repository.
\end{enumerate}

\section{Preliminary}\label{sec:backgrouds}
Here we briefly recap QNNs that will be explored in this study. The foundation of quantum computing is provided in Appendix \ref{subsec:prep-Qc}. Please refer to literature \cite{benedetti2019parameterized,cerezo2020variational2,nielsen2010quantum,wittek2014quantum} for comprehensive explanations.

\subsection{Quantum neural network}\label{subsec:prep-QNN}
 
\begin{figure*}[htp]
	\centering \includegraphics[width=0.9\textwidth]{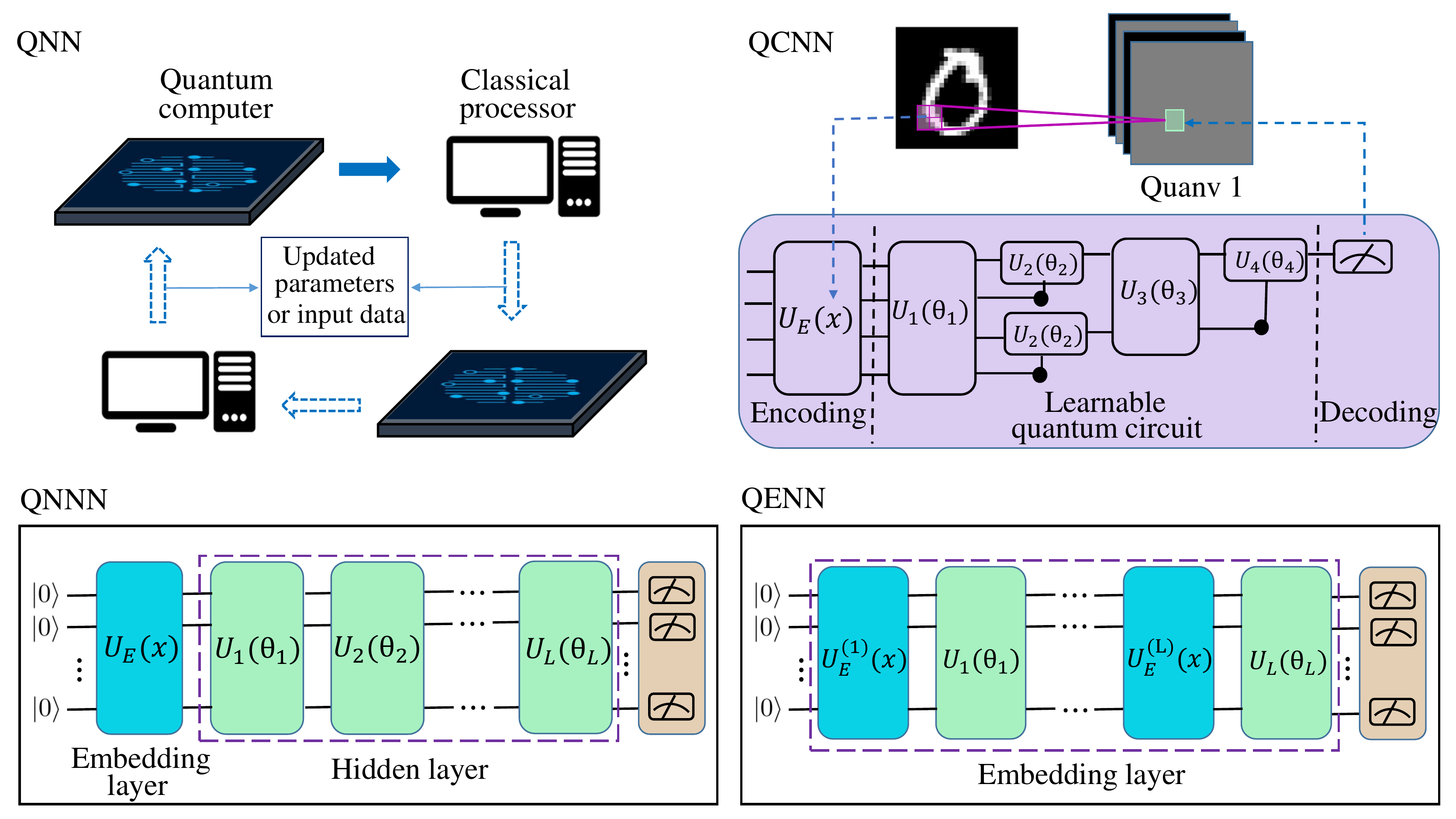}
	\caption{\small{\textbf{The machinery  of various QNNs}. The schematic of QNN, depicted in the upper left, consists of a hybrid quantum-classical loop, where the quantum computer is employed to train the learnable parameters and the classical processor is utilized to perform the optimization or post-processing to the collected information from the quantum computer. The dashed arrows mean that the loop is finite and terminated in the classical processor. The paradigms of QNNN, QENN, and QCNN are shown in the lower left, lower right, and upper right, respectively. The detailed realization of these QNNs is presented in Sections \ref{subsec:QNNN}, \ref{subsec:QENN}, and \ref{subsec:QCNN}. }}
	\label{fig:QNN_schem}
\end{figure*}

Quantum neural networks (QNNs) can be treated as the quantum generalization of deep neural networks (DNNs), as illustrated in Figure \ref{fig:QNN_schem}. Both of them leverage an optimizer to iteratively update parameters $\bm{\theta}$ of a trainable model $h(\bm{\theta}, \cdot)$ to minimize a predefined loss function $C(\cdot, \cdot)$.

The key difference between QNNs and DNNs is the strategy to implement the trainable model $h(\bm{\theta}, \cdot)$, where the former employs the parameter quantum circuits (PQCs), or equivalently ans\"atzes \cite{benedetti2019parameterized,cerezo2020variational2,du2018expressive}, while the latter utilizes neural networks \cite{goodfellow2016deep}. In particular, PQCs are constituted by the encoding part $U_E(\cdot)$, the trainable part $U(\bm{\theta})$, and the readout part. The purpose of $U_E(\cdot)$ is loading classical information into the quantum form, as the precondition to proceed further quantum operators. Although there are many data encoding methods \cite{larose2020robust}, here we mainly focus on the qubit-encoding method and its variants, because of their resource-efficient property. Once the quantum example is prepared, the trainable unitary $U(\bm{\theta})$ is applied to this state, followed by the quantum measurement $\{\Pi_i\}$ to extract quantum information into the classical form (see following subsections for details). The collect classical information can either be used as the predicted label or the hidden feature, depending on the detailed QNN-based protocols.  

In the subsequent three subsections, we   elaborate on the implementation of three representative protocols, i.e., quantum naive neural network (QNNN) \cite{havlivcek2019supervised}, quantum embedding neural network (QENN) \cite{lloyd2020quantum}, and quantum convolutional neural network (QCNN) \cite{henderson2020quanvolutional}, respectively.  

\subsubsection{Quantum naive neural network}\label{subsec:QNNN}
We first follow Eqn.~(\ref{eqn:QNN-overview}) to elaborate on the implementation of PQCs, or equivalently, the hypothesis $h(\bm{\theta}, \bm{x}^{(i)})$, in QNNN. As shown in Figure~\ref{fig:QNN_schem}, the encoding circuit $U_E(\cdot)$ loads the classical example into the quantum state by specifying data features as rotational angles of single-qubit gates. Note that the topology of quantum in $U_E(\cdot)$ can be varied, e.g., a possible implementation is $U_E(\bm{x}^{(i)})=\bigotimes_{j=1}^d \RY(\bm{x}^{(i)}_j)$ \cite{lloyd2020quantum}. The trainable part $U(\bm{\theta})$ consists of trainable single-qubit quantum gates and fixed two quantum gates. Analogous to $U_E(\cdot)$, the topology of $U(\bm{\theta})$ are versatile, where involving more gates promises a higher expressivity but a more challenged trainability \cite{du2021efficient,holmes2021connecting}. Here we mainly focus on the hardware-efficient structure such that the construction of $U(\bm{\theta})$ obeys a layer-wise structure and the gates arrangement in each layer is identical. The explicit form satisfies
	$U(\bm{\theta}) = \prod_{l=1}^L U_l{(\bm{\theta}_{l})}$, where $L$ is the layer number and $\bm{\theta}_{l}$ denotes the trainable parameters at the $l$-th layer.  To extract the quantum information into the classical form, QNNN applies POVMs $\{\Pi_i\}_{i=1}^{d_y}$ to the state $\ket{\psi(\bm{x}^{(i)},\bm{\theta})}=U(\bm{\theta}) U_E(\bm{x}^{(i)}) \ket{0}^{\otimes d}$, i.e.,
	\begin{eqnarray}\label{eqn:QNNN-hypothesis}
		h(\bm{\theta}, \bm{x}^{(i)}) =    \Big[ & \Tr\left(\Pi_1  \ket{\psi(\bm{x}^{(i)},\bm{\theta})}\bra{\psi(\bm{x}^{(i)},\bm{\theta})} \right), \cdots,  
		   \Tr\left(\Pi_{d_y}  \ket{\psi(\bm{x}^{(i)},\bm{\theta})}\bra{\psi(\bm{x}^{(i)},\bm{\theta})} \right)  \Big]^{\top}, 
	\end{eqnarray}
where $d_y$ is the dimension of the label space. In the training process, we adopt the first-order optimizer to update the parameters $\bm{\theta}$ to minimize the loss function in Eqn.~(\ref{eqn:QNN-overview}), where the gradients $\partial C\left(h(\bm{\theta}, \bm{x}^{(i)}), y^{(i)} \right)/ \partial \bm{\theta}$ can be analytically evaluated by the parameter shift rule \cite{mitarai2018quantum}. The specific optimization algorithms used in this study are stochastic gradient descent (SGD) \cite{kingma2014adam} and stochastic quantum natural gradient descent (SQNGD) \cite{stokes2020quantum}. More details can be seen in Appendix~\ref{subsec:optimization}.

\subsubsection{Quantum embedding neural network }\label{subsec:QENN}
Instead of separating the encoding part from the training part, QENN integrates them together into the embedding circuit where the encoding circuit and the training circuit are carried out alternately. Specifically, in QENN, the employed PQCs are composed of multiple embedding layers equipped with trainable parameters, i.e., in each layer an encoding circuit $U_E^{(l)}(\bm{x}^{(i)})$ is followed by a trainable circuit $U_l(\bm{\theta}_l)$, as shown in Figure~\ref{fig:QNN_schem}. Throughout the whole study, we consider an identical topology of $U_E^{(l)}(\cdot)$ for different layers, as the repetition of embedding layer is demonstrated to implement the classically intractable feature maps~\cite{lloyd2018quantum_app} and increase the expressive power of QNN~\cite{schuld2021effect}. The  explicit form of such PQCs can be written as $U(\bm{x},\bm{\theta})=\prod_{l=1}^{L}U_E(\bm{x}^{(i)})U_{l}(\bm{\theta}_l)$, where the meanings of $L$ and $\bm{\theta}_l$ are the same with those in QNNN. As for the training of QENN, the strategy of measurement and optimization are identical to QNNN in Subsection \ref{subsec:QNNN}.

\subsubsection{Quantum convolutional neural network }\label{subsec:QCNN}
Convolutional neural network (CNN) has demonstrated the superiority in images processing tasks, including two special local operators, i.e., convolution and pooling. The function of convolutional and pooling operations is extracting local features from the whole image and aggregating information from adjacent patches, respectively. Unlike CNN, quantum CNN (QCNN) \cite{henderson2020quanvolutional} completes the convolutional operation by the quantum convolutional layer to pursue better learning performance. Particularly, in the quantum convolutional layer, a fraction of the input image is embedded into the quantum circuit $U_E(\cdot)$, interacted with PQCs $U(\bm{\theta})$, followed by the quantum measurements to extract the corresponding semantic features. As shown in Figure \ref{fig:QNN_schem}, unlike QNNN and QENN, where the collected classical information is directly utilized as the predictable label, the output of the quantum convolutional layer is treated as a hidden feature map which is taken as the input for the next quantum convolutional layer. After several quantum convolutional operations, a classical fully connected layer with activation function \cite{goodfellow2016deep} acts on the extracted features to make predictions.

\section{The generalization of quantum neural networks}\label{sec:generalization}
In this section, we explore the generalization ability of representative QNNs introduced in Section \ref{sec:backgrouds}. To be more concrete, we first apply these QNNs to learn  real-world datasets and compare their performance with classical DNNs with varied parameter settings, i.e., the multi-layer perceptron (MLP) and  convolutional neural network (CNN) whose number of trainable parameters is similar to QNNs, the over-parameterized multi-layer perceptron (MLP++) whose number of trainable parameters can be extremely large \cite{he2016deep,krizhevsky2012imagenet}. We further conduct systematical simulations to benchmark  the \textit{effective model capacity} of QNNs and DNNs, since this measure the determines  generalization ability of learning models \cite{mohri2012foundations,zhang2017rethink},

We experiment on two real-world datasets, i.e., the Wine dataset \cite{Dua:2019} and MNIST handwritten digit dataset \cite{lecun1998mnist}, to examine the generalization ability of QNNs and DNNs. The Wine dataset, collected from UCI Machine Learning Repository \cite{Dua:2019}, consists of 130 examples, where each example is described by a feature vector with $13$ attributes determining the origin of wines. MNIST dataset includes ten thousands hand-written digit images, where each image has $28\times 28$ pixels. With the aim of removing data-dependent bias as much as possible, we also assess the generalization ability of the quantum synthetic data proposed by \cite{havlivcek2019supervised}. Specifically, we train QNNN, QENN, and MLP on the Wine dataset and quantum synthetic dataset which represent 1-dimensional features; and apply QCNN, CNN and MLP on MNIST dataset for the case of $2$-dimensional features. Note that QNNN and QENN are excluded when processing image data because it requires unaffordable number of qubits when embedding the high dimensional image into a quantum circuit. For suppressing the effects of randomness, the statistical results are collected by repeating each setting with $10$ times.   
  
\begin{figure*}[htp]
    \centering
    \includegraphics[scale=0.5]{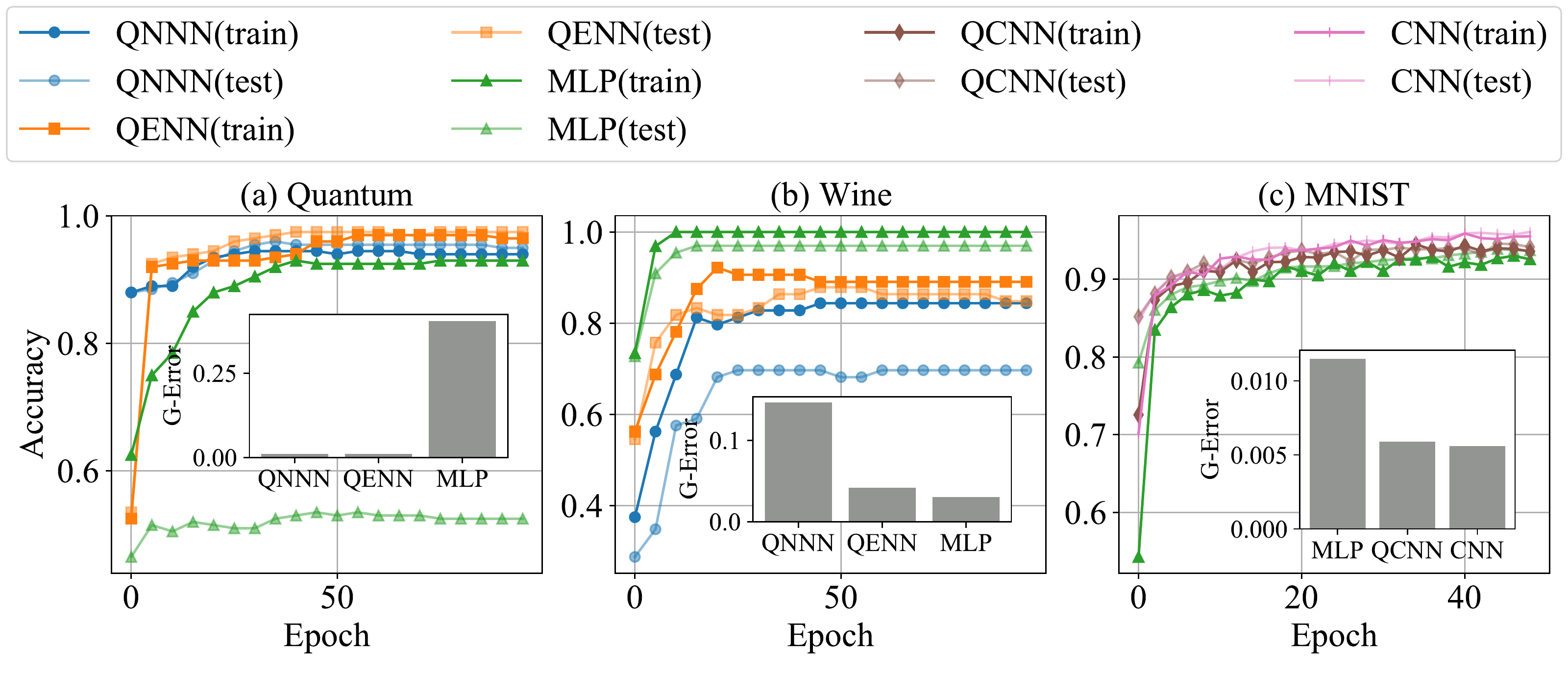}
    \caption{\small{\textbf{Learning performance on quantum data and classical data with \textit{true labels}}. \textsf{G-Error} represents the generalization error. (a), (b) and (c) show the accuracy of various models changing with training epochs, when training on quantum synthetic data, the Wine data and MNIST respectively. The bar chart inserted into each figure represents the generalization error of each model.}}
    \label{fig:performance}
\end{figure*}

Before moving on to present experiment results, let us address the  generalization error measure defined in Eqn.~(\ref{eqn:gene-error}). Particularly, there are two components that together completely characterize the generalization error, i.e., the empirical risk $\hat{\mathcal{R}}_S(\hat{\bm{\theta}})$ and the expected risk $\mathcal{R}(\hat{\bm{\theta}})$. In our experiments, we employ the accuracy of the training set to quantify the  empirical risk in which high train accuracy reflects low $\hat{\mathcal{R}}_S(\hat{\bm{\theta}})$. Meanwhile, following the explanation in Eqn.~(\ref{eqn:gene-error}), the accuracy on the test set is adopted to estimate the expected risk such that high test accuracy implies low $\mathcal{R}(\hat{\bm{\theta}})$. Under the above insights, when a learning model possesses a good generalization ability, it should achieve high train accuracy and test accuracy, as well as a small gap between them.

The learning performance of QNNs and DNNs for the quantum synthetic dataset and real-world datasets is exhibited in  Figure \ref{fig:performance}. Towards the quantum synthetic dataset, both the train and test accuracy of QNNN and QENN fast converge to the $92\%$ after $20$ epoch. Conversely, although the train accuracy of MLP reaches $90\%$ after $30$ epoch, its test accuracy is no better than the random guess. Therefore, the generalization error of classical DNNs, i.e., the discrepancy between train accuracy and test accuracy, is much higher than that of quantum models ($0.4$ for MLP versus $0.01$ for QNNN and QENN), as demonstrated in the bar chart inserted in Figure \ref{fig:performance} (a). However, the learning performance behaves quite different when the above models are applied to learn real-world datasets. As shown in Figure \ref{fig:performance} (b), there exists an evident step-by-step accuracy dropping on Wine dataset along the sequence of MLP, QENN, and QNNN. In particular, QNNN and QENN fall behind MLP by $10\%$ to $20\%$. Meantime, there is a more serious performance degradation for quantum models evaluated by test accuracy, especially for QNNN which holds almost $15\%$ generalization error that is three times higher than that of MLP. The learning performance of QCNN and CNN on MNIST dataset obeys the same manner. As depicted in Figure \ref{fig:performance} (c), QCNN achieves $93\%$ accuracy on both training and test set, which is slightly worse than CNN by approximately $3\%$. It is worth noting that the relatively small gap among three models is most attributed to the subtle differences in the network structure, where QCNN, CNN and MLP only differs in the first layer (Appendix \ref{append:experiment}).

\begin{figure*}[htp]
    \centering
    \includegraphics[scale=0.5]{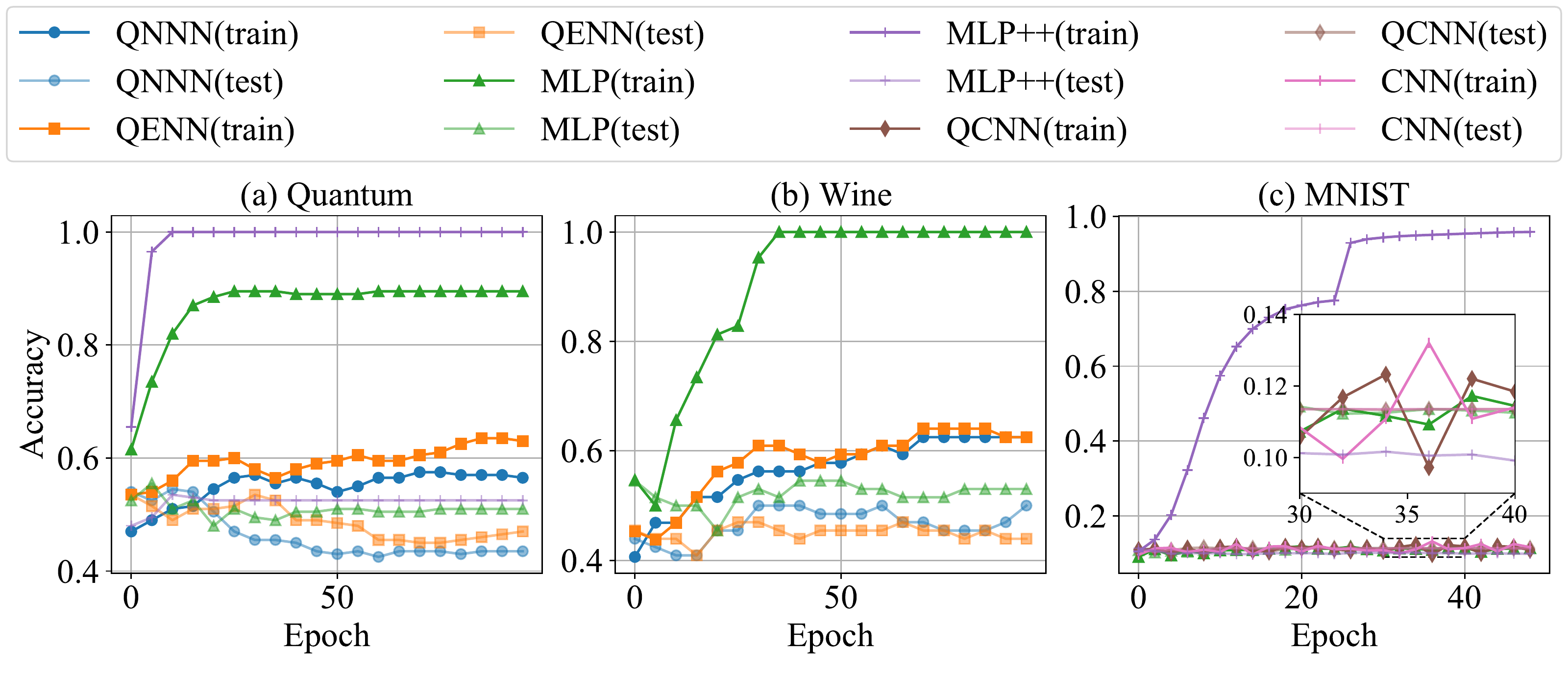}
    \caption{\small{\textbf{Trainability of different models on quantum data and classical data with \textit{random labels}}. (a) shows how various models fit quantum data with random labels. \textsf{MLP++}, representing MLP with larger scale, achieves zero training error. (b) shows the changes of accuracy when fitting classical data with random labels. MLP can still completely fit the random labels. (c) shows the ability of fitting MNIST with random labels. No model performs better than random guess.}}
    \label{fig:performance:random}
\end{figure*}

The generalization ability of a learning model is dominated by its effective model capacity, which concerns the model's ability to fit random labels \cite{zhang2017rethink}. Namely, a learning model possesses a high effective model capacity when it reaches a high train accuracy on a dataset with random labels, as ensured by the randomization test in non-parametric statistics. Empirical studies have validated that DNNs possess sufficient effective model capacity and speculated that this property contributes to the great success of DNN. With this regard, we conduct randomization experiments on both QNNs and DNNs to compare their effective model capacity.

The results related to the effective model capacity are shown in Figure \ref{fig:performance:random}, which reveal a large gap between QNNs and DNNs, regardless of whether the training data is quantum or classical. In particular, QNNs achieve relatively low train accuracy ($0.562$ for QNNN and $0.630$ for QENN) which is only slightly better than random guess. By contrast, for DNNs, MLP reaches $90\%$ and the perfect $100\%$ train accuracy on the quantum synthetic dataset and Wine dataset respectively after $40$ epochs. If we further increase the number of trainable parameters of MLP (MLP++), it will also completely fit the quantum synthetic data with random labels, as shown by the purple line in Figure \ref{fig:performance:random} (a). Note that the same strategy is inappropriate to be applied to QNNs, because increasing trainable parameters will incur both the barren plateau issues and the accumulated quantum noise, which hinder the optimization \cite{du2020learnability}. As for MNIST, all these models fail to fit the random labels, whose behavior imitates the random guess. Similarly, we can enlarge the scale of MLP to obtain a leap of accuracy on training set with random labels at the expense of little growth of training time, as indicated by the purple line in Figure \ref{fig:performance:random} (c).

\textbf{Remark.} We defer the study of the generalization ability of QNNs under the NISQ case in Appendix \ref{append:result}. In a nutshell, by realizing QNNs on NISQ chips to complete above experiments, we conclude that \textit{quantum noise largely weakens the effective model capacity and generalization of QNNs}. 

\subsection{Implications}
The achieved results indicate the following three substantial implications with respect to the dilemma of existing QNNs. 
\begin{enumerate}
	\item The learning performance of current QNNs is no better than DNNs on real-world datasets. This observation questions the necessity to employ QNNs to tackle real-world learning tasks, since it remains elusive how QNNs can benefit these tasks. 
	\item The effective model complexity of current QNNs is poor, which is stark contrast with DNNs. The low effective model complexity enables us to leverage statistical learning theory to analyze the generalization ability of QNNs with a tight bound \cite{huang2020power,banchi2021generalization}.  Nevertheless, as shown in Figure \ref{fig:schem}, a severely restricted model capacity  fails to cover complicated target concepts in real-world tasks, which prohibits the applicability of QNNs. 
	\item The limited model capacity is further reduced by imperfection of NISQ machines. The narrowed  hypothesis space deteriorates the performance of QNNs. 
\end{enumerate}
 
There are two possible directions to seek potential advantages of QNNs over DNNs. The first way is designing clever over-parameterization quantum learning models as with DNNs. Partial evidence to support this solution is the improved performance of QENN compared with QNNN. A critical issue in such a model design is how to avoid barren plateau phenomenons \cite{mcclean2018barren}.  The second way is to develop new paradigm of quantum models to further introduce nonlinearity into quantum learning pipeline. For instance, theoretical results have proven potential advantages of quantum kernels \cite{havlivcek2019supervised,schuld2021quantum,wang2021towards}.

\section{Trainability of quantum models}\label{sec:trainability}
Here we investigate the trainability of QNNs, which serves as another dominant factor manipulating the learnability of quantum models. In particular, we first examine the performance of QNNN and QENN on the Wine dataset with consideration of various implicit and explicit regularization techniques under the noiseless scenario. Subsequently, for the purpose of understanding how quantum noise affects the performance of QNNs, we conduct the same experiments under the noisy scenario in which the noisy model is extracted from ibmq\_16\_melbourne, which is one of the IBM Quantum Canary Processors \cite{IBMQ}.

Recall a learning model, e.g., QNNs or DNNs, is warranted to possessing a good trainability if it requires a small number of epochs to surpass a threshold accuracy and fast converges to a stationary point, i.e., the term $\mathcal{J}(\bm{\theta})\rightarrow 0$ in   Eqn.~(\ref{eqn:trainability}). Many theoretical studies in the regime of deep learning have proven that SGD can help learning models escape saddle points efficiently \cite{bottou2018optimization,ge2015escaping}. Moreover, recent study demonstrated that a quantum variant of SGD, i.e., stochastic quantum natural gradient descent (SQNGD) \cite{stokes2020quantum}, can well address the barren plateau's issue. Driven by the success of SGD in deep learning and the power of SQNGD, an immediate interest is exploring whether these two methods can enhance the trainability of QNNs, especially for alleviating barren plateaus issues.  
 
 \begin{figure*}[htp]
    \centering
    \subfigure[]{
    \includegraphics[scale=0.38]{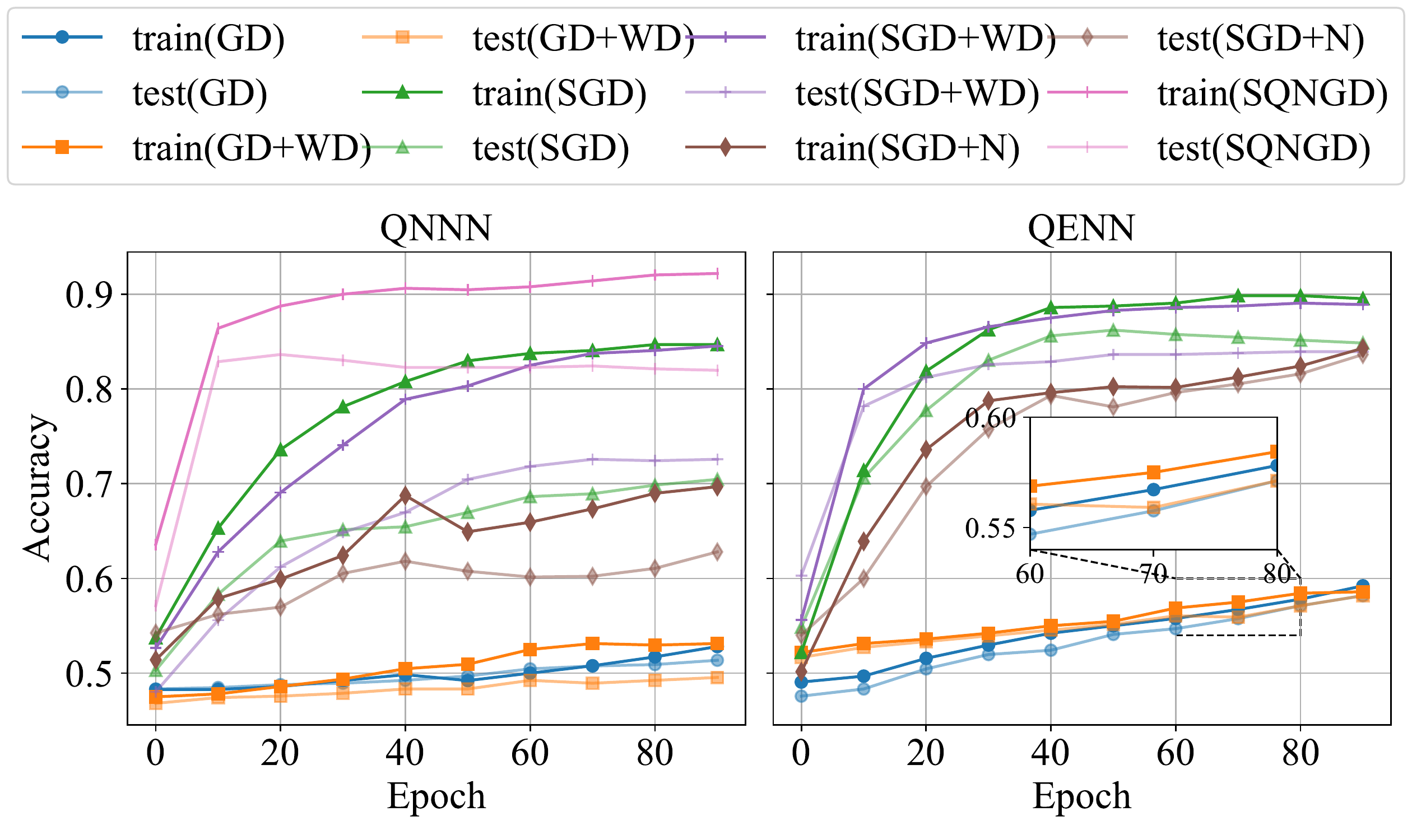}}
    \subfigure[]{
    \includegraphics[scale=0.38]{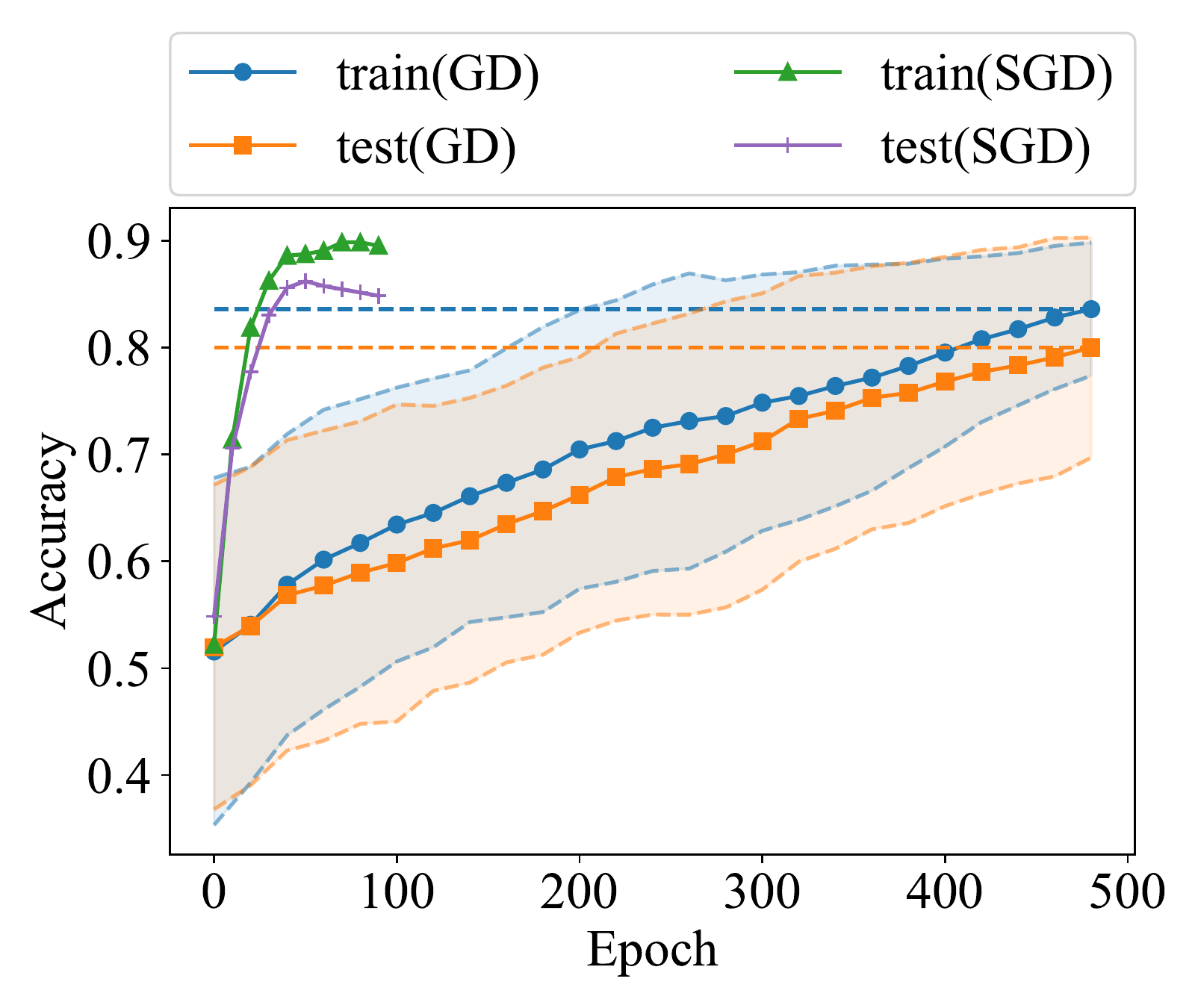}}
    \caption{\small{\textbf{Effects of regularizations on the performance of quantum model on  Wine dataset}. The labels  `\textsf{GD}', `\textsf{SGD}', `\textsf{SQNGD}', `\textsf{WD}', and  `\textsf{N}' refer to the   gradient descent optimizer,  stochastic gradient descent optimizer,  the stochastic quantum natural gradient descent optimizer, the weight decay, and execution of experiments on NISQ chips, respectively. (a) describes the effects of regularizations on optimization. SGD plays a significant role in accelerating convergence and achieving higher accuracy, while others the optimization process instead of boosting performance. (b) is the learning curve of GD with more training epochs.}}
    \label{fig:general:wine}
\end{figure*}

The experiment results are summarized in  Figure \ref{fig:general:wine}. With the aim of investigating whether SGD facilitates the quantum model optimization, we also apply gradient descent (GD) optimizer to learn the same tasks as a reference. Specifically,  Figure \ref{fig:general:wine} (a) depicts that the train accuracy of QNNN and QENN optimized by SGD rapidly rises to $70\%$ after $20$ epochs, while its train accuracy remains at around $50\%$ with the GD optimizer. Meanwhile, QNNN with SGD optimizer  achieves higher test accuracy than that in the setting of GD (at least $20\%$). Surprisingly, SQNGD, the quantum natural gradient version of SGD, further expands the accuracy gap by $10\%$, reaching to the highest accuracy  and fastest convergence of QNNN. Notably, the performance of QNNN with GD optimizer presents a monotone increasing trend, which suggests that the model may potentially reach higher performance with more training epochs. Therefore, we extend the total training epochs from $100$ to $500$ and train the QENN with GD. Figure \ref{fig:general:wine} (b) shows that the accuracy of QENN trained by GD has been increasing smoothly and reaches $80\%$ after $500$ epochs, narrowing the gap between GD and SGD from $30\%$ to $10\%$.

\begin{figure*}[htp]
    \centering
    \subfigure[Training process]{
    \includegraphics[scale=0.36]{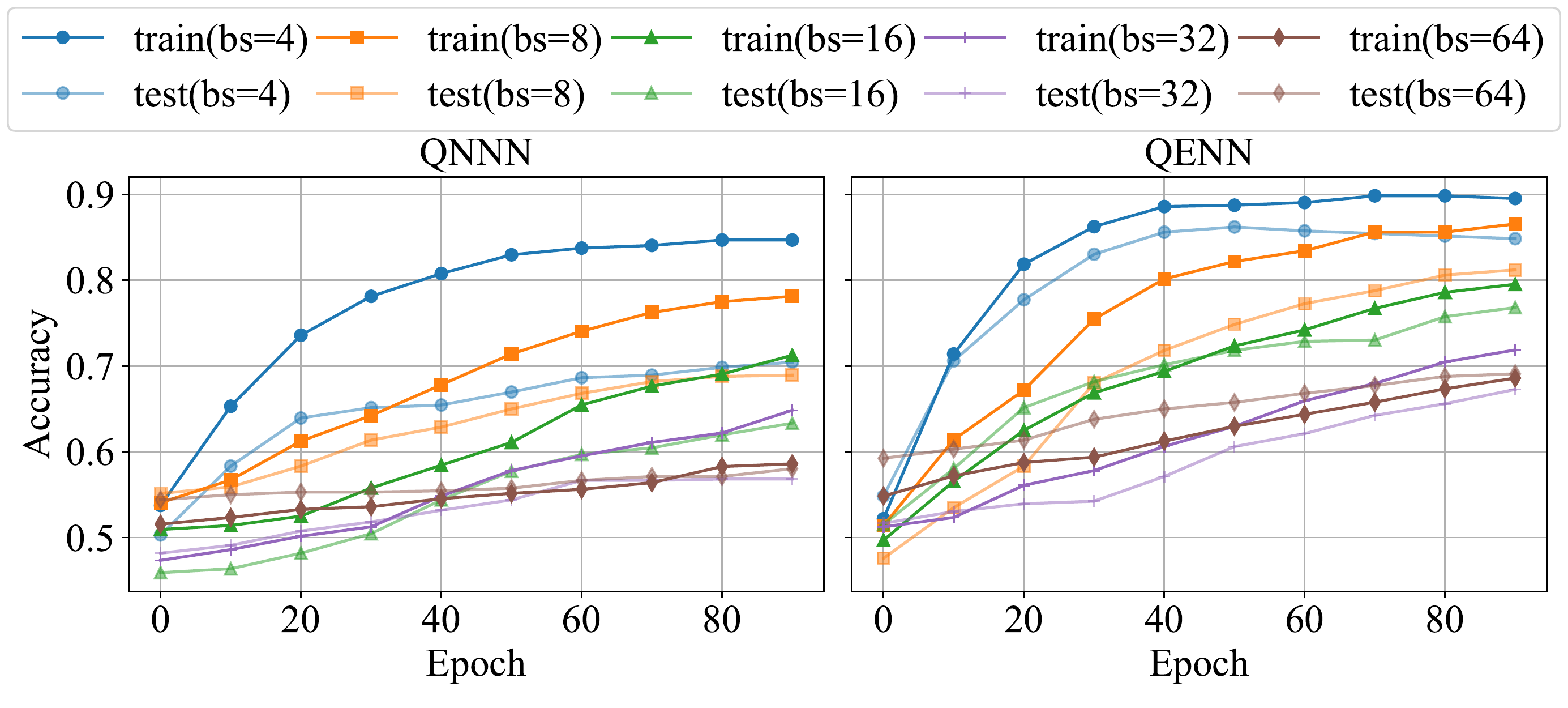}}
    \subfigure[Accuracy vs batch size]{
    \includegraphics[scale=0.36]{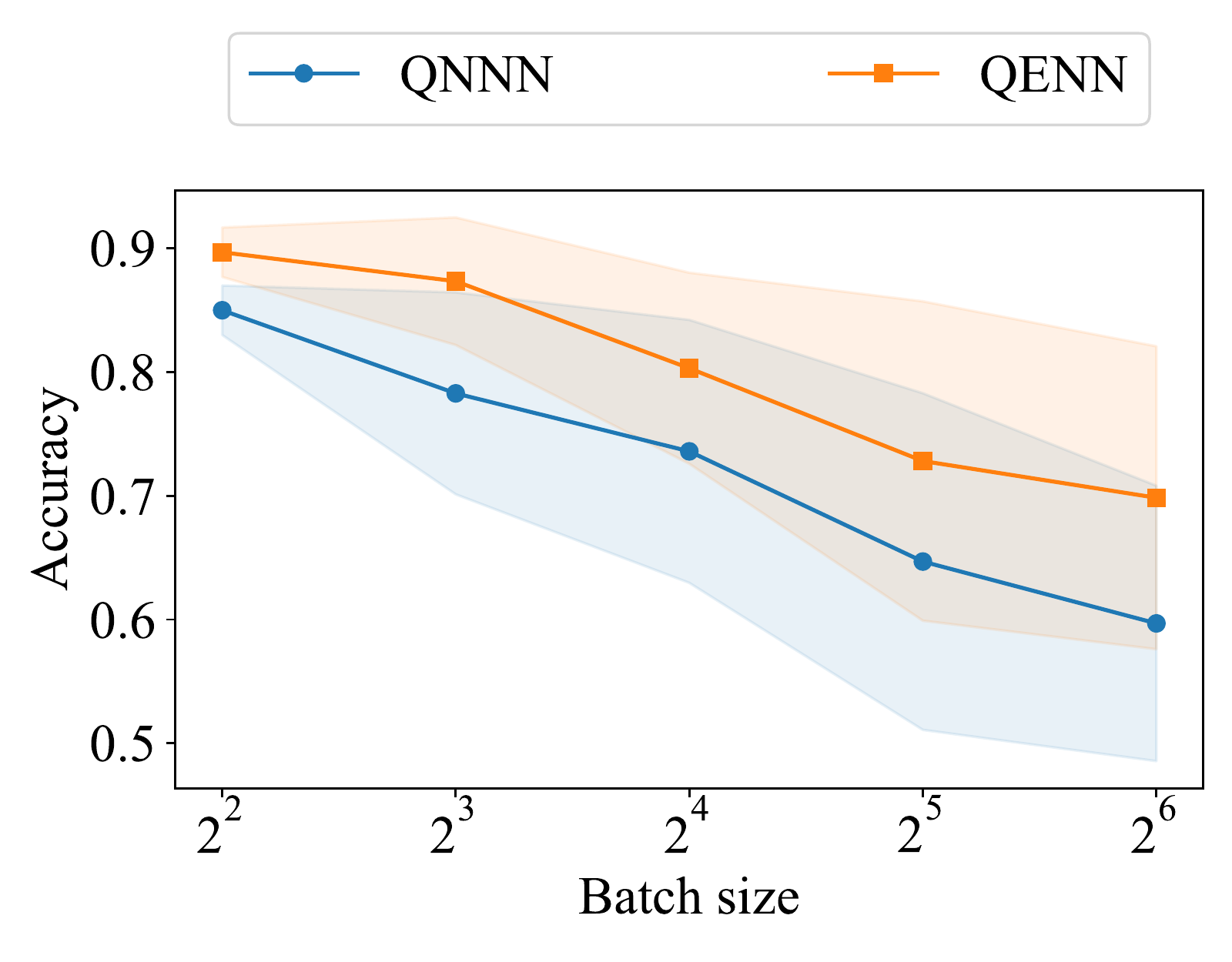}}
    \caption{\textbf{The relation between batch size of SGD and trainability of QNNs on Wine dataset}. (a) shows the convergence of quantum models with different setting of batch size. \textsf{bs} is abbreviation of batch size. (b) shows the the fluctuation of training accuracy with respective to batch size.}
    \label{fig:batchsize:wine}
\end{figure*}

Motivated by the large gain from SGD, we conduct additional experiments to exploit how the batch size affects the learning performance of QNNs. Specifically, we train QNNs many times on the Wine data by SGD with batch size growing exponentially. As shown in Figure \ref{fig:batchsize:wine}, for both QNNN and QENN, the increased batch size suppresses the convergence rate. For example, QNNN achieves $80\%$ accuracy on the Wine dataset when batch size equals to $4$ after $40$ epochs, which is $10\%$ higher than that of batch size with $8$.

We then study how the explicit regularization technique, i.e., the weight decay, effects the trainability of QNNs.  Mathematically, the weight decay stragety refers to adding a penalty term for the trainable paremters, i.e., $\arg \min_{\bm{\theta}} \mathcal{L}(\bm{\theta})=\frac{1}{n}\sum_{i=1}^n\ell(y^{(i)}, \hat{y}^{(i)})+\lambda\|\bm{\theta}\|$. The effect of weight decay on the trainability of QNNs is shown in Figure \ref{fig:general:wine}. An immediate observation is that this strategy fails to enhance the performance of QNNN with GD optimizer. With respect to SGD optimizer, the weight decay method improves the test accuracy of QNNN by $2\%$ at the expense of slightly low convergence rate. For QENN, weight decay together with SGD assists models to obtain the fastest convergence in the first $20$ epochs. However, it fails to efficiently narrow the gap between train accuracy and test accuracy after the $40$-th epoch when the phenomenon of over fitting begins to appear.

We last explore the performance of QNNs under the NISQ scenario. As depicted by the brown line in Figure \ref{fig:general:wine},  quantum noise leads to the degraded trainability QNNs, i.e., around $10\%$ accuracy decline compared with noiseless settings. We defer the comparison of the runtime cost of simulating QNNs and DNNs in Appendix \ref{append:result}, which shows that QNNN under NISQ setting spends up to $126s$ on one iteration, while the noiseless setting and classical MLP only need $4s$ and $0.02s$ respectively.

\subsection{Implications}

The achieved results indicate that the widely used regularization techniques in classical deep learning plays a different role in quantum machine learning. Although SGD with the appropriate batch size slightly benefits the optimization of QNNs, others regularization strategies such as weight decay fail to enhance the trainability of QNNs. This differs QNNs with DNNs. Advanced techniques are highly desired to improve the trainability of QNNs, especially addressing the barren plateaus phenomenons. In addition, empirical results exhibit that quantum system noise exacerbates the training difficulty of QNNs. A promising way to resolve this issue is introducing various error mitigation techniques into QNNs \cite{du2020quantumQAS,endo2018practical,endo2021hybrid,kandala2019error,strikis2020learning,temme2017error}.

\section{Discussion and conclusions}\label{sec:conclusion}

In this study, we proceed systematic numerical experiments to understand the generalization ability and trainability of typical QNNs in the view of statistical learning theory. The achieved results exhibited that current QNNs struggle a poor effective model capacity. As depicted in Figure \ref{fig:schem}, this observation well explains why current QNNs can attain computational advantages on quantum synthetic data classification tasks and discrete logarithm problems, while they  fail to compete with DNNs in tackling real-world learning tasks. Moreover, our study illustrate that the regularization techniques, which greatly contribute to the success of DNNs, have limited effects towards the trainability of QNNs. In addition, our study exhibits that quantum system noise suppresses the  learnability of QNNs, which echoes with the theoretical study \cite{du2021efficient}. Last, to alleviate the dilemma of current QNNs, we discuss several prospective directions such as designing over-parameterized QNNs without barren plateaus and developing effective error mitigation techniques.

Besides the contributions towards the understanding the power of QNNs, we build an open-source benchmark to fairly and comprehensively assess the learnability of various QNNs in a standard process, and consequently benefit the design of new paradigms of QNNs. Specifically, this benchmark provides several ready-to-use datasets, quantum and classical models as well as evaluation scripts. Furthermore, we adopt the factory method in the software design to help users easily register their self-defined models into the whole framework. More models and tasks will be supported in the future. We believe that this benchmark will facilitate the whole quantum machine learning community.
  
\textbf{Note added.} During the preparation of the manuscript, we notice that a very recent theoretical study \cite{caro2021encodingdependent} indicated that to deeply understand the power of QNNs, it is necessary to demonstrate whether QNNs possess the ability to achieve zero risk for a randomly-relabeled real-world classification task. Their motivation highly echoes with our purpose such that statistical learning theory can be harnessed as a powerful tool to study the capability and limitations of QNNs. In this perspective, the achieved results in this study provide a negative response to their question. Combining the analysis in \cite{caro2021encodingdependent} and our results, a promising research direction is analyzing the non-uniform generalization bounds of QNNs to understand their power.


\newpage   
\clearpage 
\appendix

\section{Quantum computing}
 \label{subsec:prep-Qc}
Analogous to the fundamental role of \textit{bit} in classical computing, the fundamental unit in quantum computation is quantum bit (\textit{qubit}), which refers to a two-dimensional vector. Under Dirac notation, a qubit state is defined as  $\ket{\bm{\alpha}}=a_0\ket{0}+a_1\ket{1}\in\mathbb{C}^2$, where $\ket{0}=[1,0]^{\top}$ and $\ket{1}=[0,1]^{\top}$ specify two unit bases, and the coefficients $a_0, a_1 \in \mathbb{C}$ satisfy $|a_0|^2+|a_1|^2=1$. Similarly, an $N$-qubit state is denoted by $ \ket{\Psi}=\sum_{i=1}^{2^N}a_i\ket{\bm{e}_i} \in \mathbb{C}^{2^N}$, where $\ket{\bm{e}_i}\in\mathbb{R}^{2^N}$ is the unit vector whose $i$-th entry being 1 and other entries are 0, and $\sum_{i=0}^{2^N-1}|a_i|^2=1$ with $a_i\in\mathbb{C}$. Apart from Dirac notation, the density matrix can be used to describe more general qubit states. For example, the density matrix corresponding to the state $\ket{\Psi}$ is $\rho=\ket{\Psi}\bra{\Psi}\in \mathbb{C}^{2^N\times 2^N}$. For a set of qubit states $\{p_i, \ket{\Psi_i} \}_{i=1}^m$ with $p_i>0$, $\sum_{i=1}^{m}p_i=1$, and $\ket{\Psi_i}\in \mathbb{C}^{2^N}$ for $\forall i\in m$, its density matrix is $\rho=\sum_{i=1}^{m} p_i\rho_i$ with $\rho_i=\ket{\psi_i}\bra{\psi_i}$ and $\Tr({\rho})=1$. 

\begin{figure*}[htp]
	\centering
	\includegraphics[width=0.995\textwidth]{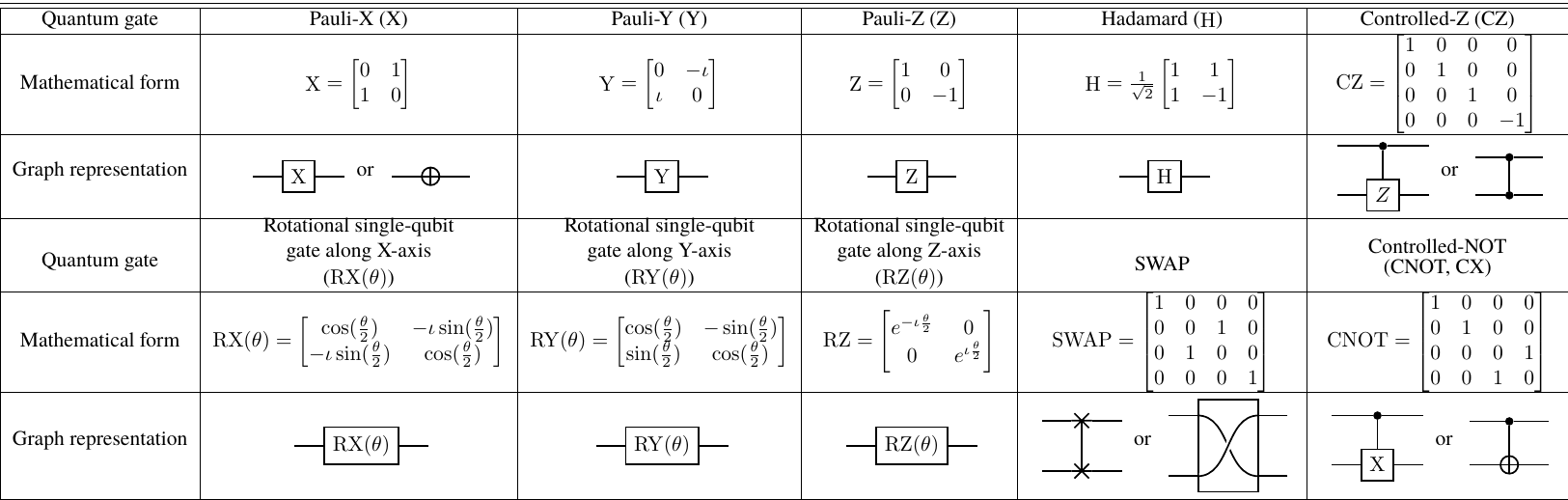}
	\caption{\small{\textbf{The quantum logic gates.} The table contains the abbreviation, the mathematical form, and the graph representation of a set of universal quantum gates explored in this study.  }}
\label{tab:Q-gates}
\end{figure*}

There are three types of quantum operations using to manipulate qubit states, which are \textit{quantum (logic) gates}, \textit{quantum channels}, and \textit{quantum measurements}.  Specifically, quantum gates, as unitary transformations, can be treated as the computational toolkit for quantum circuit models, i.e., an $N$-qubit gate $U\in\mathcal{U}(2^N)$ obeys $UU^{\dagger}=\mathbb{I}_{2^N}$, where $\mathcal{U}(\cdot)$ stands for the unitary group. Throughout the whole study, we focus on the single-qubit and two-qubit quantum gate set $\{\Hada, \XGate, \YGate, \ZGate, \RX, \RY, \RZ, \CNOT, \CZ \}$, as summarized in Figure \ref{tab:Q-gates}. Note that the investigated set is universal such that such that these quantum gates can be used to reproduce the functions of all the other quantum gates \cite{nielsen2010quantum}. Different from quantum gates that evolve qubit states in the closed system, quantum channels are applied to formalize the evolving of qubits states in the open system. Mathematically, every quantum channel $\mathcal{E}(\cdot)$ is a linear, completely positive, and trace-preserving map \cite{nielsen2010quantum}. A special quantum channel is called depolarization channel, which is defined as 
\begin{equation}
	\mathcal{E}_{p}(\rho)=(1-p)\rho + p\frac{\mathbb{I}_{2^N}}{2^N}.
\end{equation}
Intuitively, the depolarizing channel  considers the worst-case scenario such that the information of the input state can be entirely lost with some probability. The aim of quantum measurements is extracting quantum information of the evolved state, which contains the computation result, into the classical form. In this study, we concentrate on the positive operator-valued measures (POVM), which is described by a collection of positive operators $0\preceq \Pi_i$ satisfying $\sum_i\Pi_i=\mathbb{I}$.  Specifically, applying the measurement $\{\Pi_i\}$ to the state $\rho$, the probability of outcome $i$ is given by
\begin{equation}
	\Pr(i) = \Tr(\rho\Pi_i). 
\end{equation}

\section{Optimization of QNNs}
\label{subsec:optimization}
There are many methods to optimize Eqn.~(\ref{eqn:QNN-overview}) when the hypothesis is represented by Eqn.~(\ref{eqn:QNNN-hypothesis}), which include the zero-order, the first-order, and the second-order optimizers \cite{benedetti2019parameterized}. For clearness, here we concentrate on the  first-order optimizer, where the gradients $\partial C\left(h(\bm{\theta}, \bm{x}^{(i)}), y^{(i)} \right)/ \partial \bm{\theta}$ are obtained by the parameter shift rule \cite{mitarai2018quantum}, which can be effectively realized on NISQ devices. Specifically, we denote the parameter as $\bm{\theta}=(\bm{\theta}_1^T, \cdots, \bm{\theta}_L^T)^T$ with $\bm{\theta}_l=(\theta_{l1}, \cdots, \theta_{l, k_l})^T$, where the subscript $l_k$ refers to the number of parameters or gates of the $l$-th learnable circuit layer. Then the derivative of the hypothesis $h(\bm{\theta}, \bm{x}^{(i)})$ with respect to $\theta_{l, k_l}$ can be evaluated by running twice the same quantum circuit which differs only in a shift of this parameter. Mathematically, as each parameter gate we use in the designed circuit is generated by a Pauli operator $P_{l, k_l}$, i.e., $U_{l, k_l}(\theta_{l, k_l})=\exp(-i\theta_{l, k_l} P_{l, k_l} /2)$, the parameter shift rule yields the equality
\begin{equation}\label{eqn:para_shift}
    \frac{\partial h(\bm{\theta}, \bm{x}^{(i)})} {\partial {\theta}_{l, k_l}} =  \frac{1}{2\sin \alpha} \left[h(\bm{\theta}+\alpha \bm{e}_{l, k_l}, \bm{x}^{(i)})- h(\bm{\theta}-\alpha \bm{e}_{l, k_l}, \bm{x}^{(i)}) \right],
\end{equation}
where $\bm{e}_{l, k_l}$ is the unit vector along the $\theta_{l, k_l}$ axis and $\alpha$ can be any real number but the multiple of $\pi$ because of the diverging denominator. The gradient of $C\left(h(\bm{\theta}, \bm{x}^{(i)}), y^{(i)} \right)$ can be computed by using the chain rule. 

With the gradient $\partial C\left(h(\bm{\theta}, \bm{x}^{(i)}), y^{(i)} \right)/ \partial \bm{\theta}$ at hand, many gradient descent algorithms can be employed to find the optimal parameter point. 
In this paper, we mainly consider stochastic gradient descent (SGD) algorithm and stochastic quantum natural gradient descent (SQNGD) algorithm, which both effectively deal with the randomness of gradient induced by finite measurement and hardware noise. SGD is widely used in deep learning, whose optimization strategy is to update the trainable parameters in the steepest direction indicated by the gradient. Formally, each optimization step is given by
\begin{equation}
    \bm{\theta}_{t+1}=\bm{\theta}_{t}-\eta \nabla C\left(\bm{\theta} \right)
\end{equation}
where $\eta$ is the learning rate and $C(\bm{\theta})$ is short for the loss function regarding the parameter $\bm{\theta}$. 

Unlike SGD which chooses vanilla gradient as its optimization guidance, SQNGD employs the quantum natural gradient, a quantum analogy of natural gradient, to perform the parameter updating. While vanilla gradient descent chooses the steepest descent direction in the $l_2$ geometry of parameter space, which has been shown to be sub-optimal for the optimization of quantum variational algorithms \cite{harrow2021low}, quantum natural gradient descent works on the space of quantum states equipped with a Riemannian metric tensor (called Fubini-Study metric tensor) that measures the sensitivity of the quantum state to variations in the parameters. This method always updates each parameter with optimal step-size independent of the parameterization, achieving a faster convergence than SGD. Formally, we denote the quantum state produced by the PQC as $\ket{\psi_{\bm{\theta}}}$. The optimization rule of SQNGD involving the pseudo-inverse $g^{+}(\bm{\theta}_{t})$ of the metric tensor yields
\begin{equation}
    \bm{\theta}_{t+1} = \bm{\theta}_{t} - \eta g^{+}(\bm{\theta}_{t}) \nabla C\left(\bm{\theta} \right),
\end{equation}
where $g_{ij}(\bm{\theta}) = \Real[G_{ij}(\bm{\theta})]$ is the Fubini-Study metric tensor, and $G_{ij}(\bm{\theta})$ is the Quantum Geometric Tensor which can be written as
\begin{equation}
    G_{ij}(\bm{\theta})=\braket{\frac{\partial \psi_{\bm{\theta}} } {\partial {\theta}_i }, \frac{\partial \psi_{\bm{\theta}} } {\partial {\theta}_j } }- \braket{\frac{\partial \psi_{\bm{\theta}} } {\partial {\theta}_i }, \psi_{\bm{\theta}} } \braket{\psi_{\bm{\theta}}, \frac{\partial \psi_{\bm{\theta}} } {\partial {\theta}_j } }
\end{equation}
with $\theta_i$ being the $i$-th entry of parameter vector $\bm{\theta}$.

\section{Numerical simulation details}\label{append:experiment}
The outline of this section is as follows.
The configuration of numerical simulations conducted in Section \ref{sec:generalization} and \ref{sec:trainability} is introduced in this section. First, we will give a detailed description about the datasets used in Section \ref{sec:generalization} and \ref{sec:trainability}. Next, we present the implementation details of each type of QNNs. Last, the deployed simulation hardware and hyper-parameters settings are demonstrated.

\subsection{Dataset}

\textbf{Quantum synthetic dataset}. We randomly sample classical data from uniform distribution and then embed them into quantum circuit by assigning each classical sample to the rotation angle of each quantum rotation gate. After converting classical data $\bm{x}$ to quantum state $\rho(\bm{x})$, we run the quantum circuit shown in Figure and measure the expectation of observable $O$. The whole process is expressed as
\begin{equation}
    f_{\bm{\theta}}(\bm{x}) = \bra{\bm{0}}U^\dagger(\bm{x}, \bm{\theta})OU(\bm{x}, \bm{\theta})\ket{\bm{0}},
\end{equation}
where $U$ denotes the quantum circuit which depends on classical sample $\bm{x}$ and trainable parameter $\bm{\theta}$. For binary classification task, the label $y$ of input $\bm{x}$ is calculated by $sign(f_{\bm{\theta}}(\bm{x}))$. We totally collect 200 positive samples and 200 negative samples, and split them into training set and test set equally.

\textbf{Wine dataset}. For 1-D signal, we select the Wine Data Set from UCI Machine Learning Repository, whose feature dimension is similar to the quantum synthetic data so that they can be flexibly encoded into quantum circuit. In addition, the categories are truncated to two classes, keeping consistent with the quantum data. 

\textbf{MNIST dataset.} For 2-D images, a subset of MNIST is extracted to evaluate performance of QCNN and CNN. At the same time, the original digit images with size of $28*28$ are resized to $10*10$ for reducing resource consumption. License: Yann LeCun and Corinna Cortes hold the copyright of MNIST dataset, which is a derivative work from original NIST datasets. MNIST dataset is made available under the terms of the \href{https://creativecommons.org/licenses/by-sa/3.0/}{Creative Commons Attribution-Share Alike 3.0 license}.

In summary, the statistical characteristics of three datasets are listed in Tabel \ref{tab:dataset}.

\begin{table*}[htp]
\caption{Three datasets used in experiments}
\label{tab:dataset}
\centering
\begin{tabular}{lllll}
\toprule
Dataset          & Dimension & Class number & Training sample & Test sample \\
\midrule
Quantum synthetic data     & 16 & 2 & 200 & 200 \\
The Wine data     & 13 & 2 & 65 & 65 \\
MNIST    & $10\times10$ & 10 & 2000 & 2000 \\
\bottomrule
\end{tabular}
\end{table*}

\subsection{Implementation details}

\begin{figure*}[htp]
    \centering
    \subfigure[QNNN]{
    \includegraphics[scale=0.7]{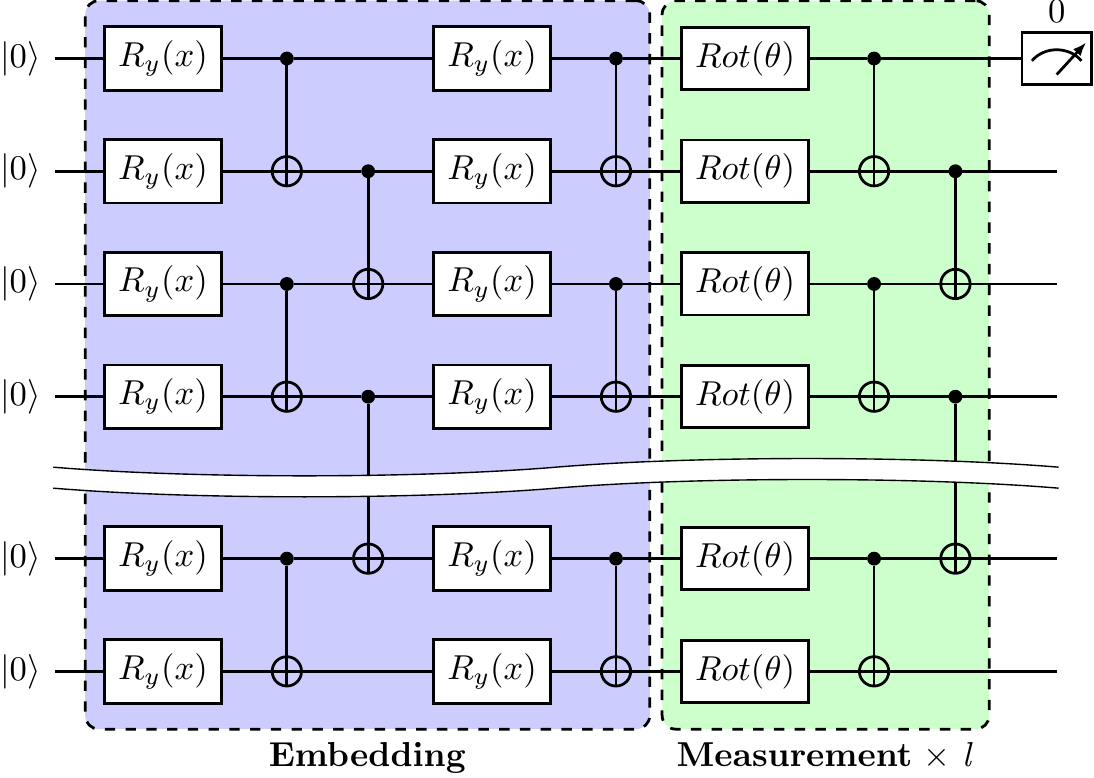}}
    \subfigure[QENN]{
    \includegraphics[scale=0.7]{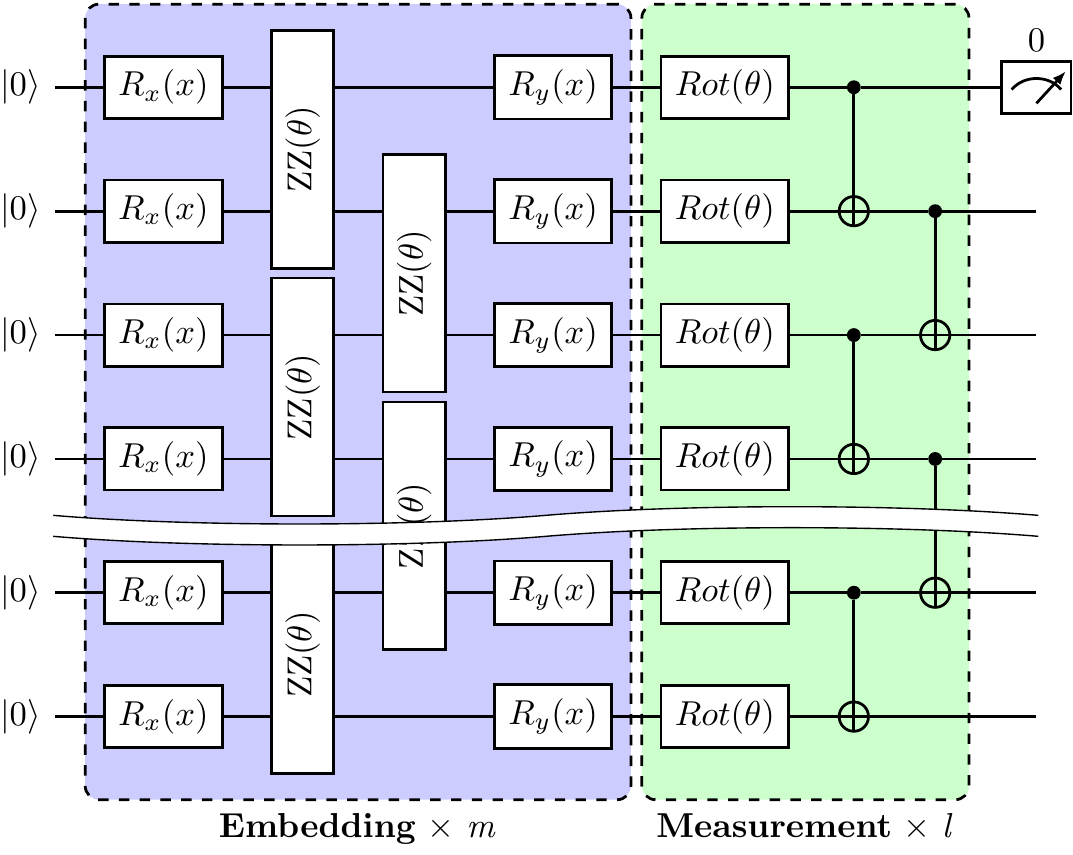}}
    \caption{\textbf{Layout of QNNN and QENN}. (a) QNNN. Gate $R_y(x)$ means the rotation gate generated by Pauli Y, whose angle is determined by the given classical data $x$. $Rot(\theta)$ represents the general rotation gate with three rotation angles $\theta_x, \theta_y, \theta_z$. $[\cdot]_{\times l}$ denotes the layer is repeated $l$ times. The initial state $\ket{0}^{\otimes n}$ is processed by the whole circuit, and the prediction of input $x$ is calculated by measuring the expectation of Pauli Z basis on the first qubit. (b) QENN. The only difference compared with the QNNN is the embedding layer, where there are additional trainable parameters to flexibly encode the classical data according to the feedback of measurement. $ZZ(\theta)$ gate is formulated by $\exp(-i \frac{\theta}{2} Z^{\otimes 2})$, which is a type of two-qubit gate introducing entanglement.}
    \label{fig:layout:QNN}
\end{figure*}

\textbf{QNNN}. QNNN is roughly divided into two blocks, including the feature embedding block and trainable measurement block. In this paper, the feature embedding block converts the classical data into quantum state by setting the parameter of quantum gates with classical data. The measurement block consists of a variational quantum circuit that linearly transforms the prepared quantum state and the measurement operation that calculates the expectation of Pauli Z basis. The number of qubits in experiments depends on the feature dimension of datasets. Concretely, we employ quantum circuits with 16, 13 and 4 qubits for quantum synthetic data, the Wine Data Set and MNIST respectively. The detailed circuit arrangement is described in Figure \ref{fig:layout:QNN} (a).

\textbf{QENN}. The basic architecture of QENN model is the same as that of QNNN. What is different is that the embedding layer also contains trainable parameters. By making the feature embedding learnable, the quantum circuit forms a more flexible and complicated transform function, which is more powerful than the vanilla QNNN. The detailed layout of QENN is shown in Figure \ref{fig:layout:QNN} (b).

\begin{figure*}[htp]
    \centering
    \includegraphics[scale=0.7]{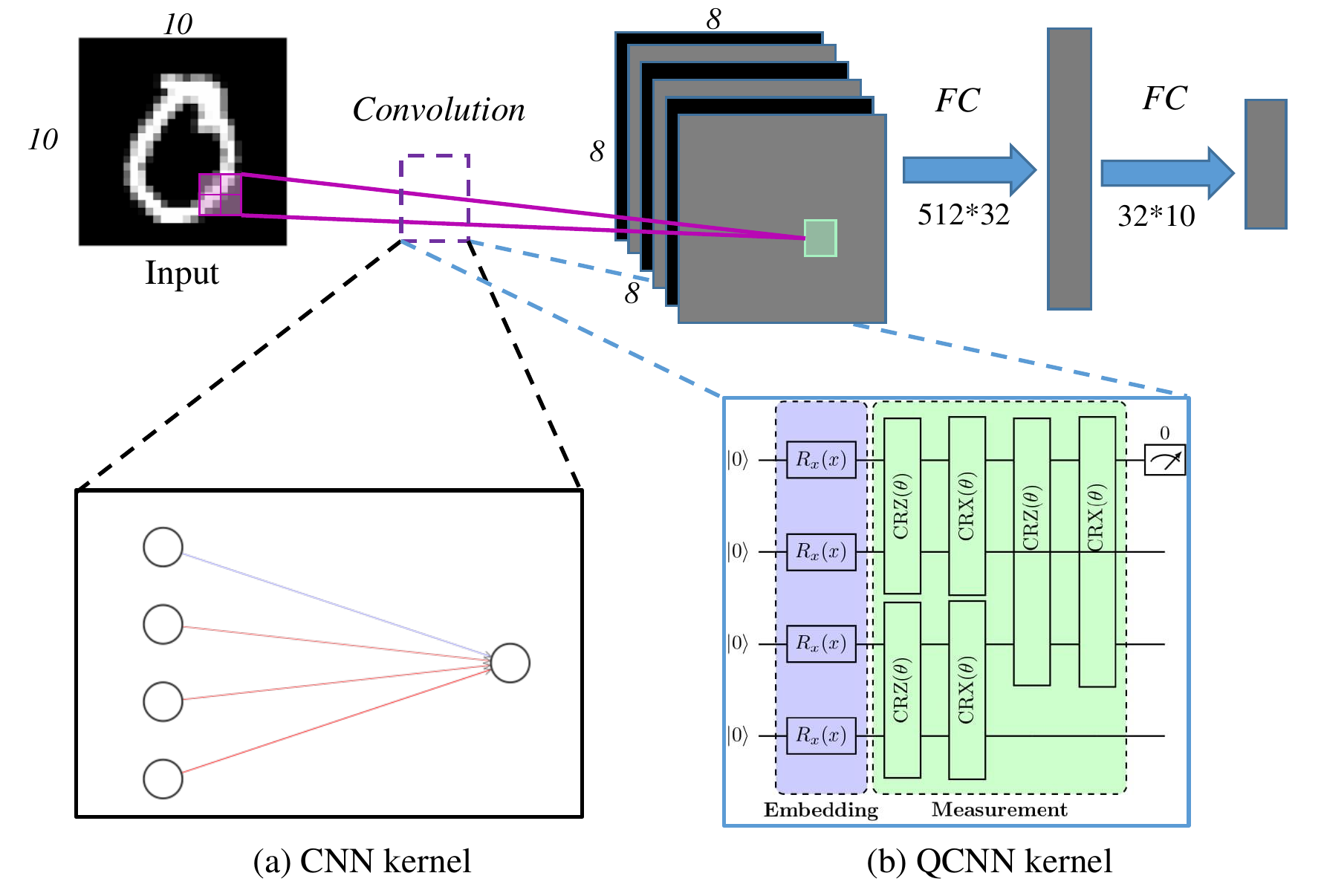}
    \caption{\textbf{Layout of CNN and QCNN}. The top half represents the overall structure of these two models. What is the difference between CNN and QCNN is the convolution kernel. (a) CNN kernel. The classical convolution kernel is implemented by the fully-connected layer, which takes the flattened pixels inside a local patch as input and outputs the convolutional result. (b) QCNN kernel. Firstly, the value of each pixel inside the same window of a kernel is encoded as the angle of $RX$ gate. Then several controlled-RZ (CRZ) and controlled-RX (CRX) operators are employed to learn the transformation. Finally, we measure the expectation of Pauli Z basis on the first qubit as the result of quantum convolution.}
    \label{fig:layout:kernel}
\end{figure*}

\textbf{QCNN}. The QCNN employed in this paper can be regarded as a quantum version of classical CNN, where the classical convolutional kernel is replaced with quantum circuit, as demonstrated in \ref{fig:QNN_schem}. In particular, we implement a $2\times2$ kernel with a variational quantum circuit of 4 qubits and 6 trainable parameters, as depicted in Figure \ref{fig:layout:kernel}. In addition, multiple duplicates of the quantum circuit are introduced to simulate the multi-channel mechanism. The feature maps returned by quantum kernels are further processed by two fully-connected layer of 32 and 10 nodes respectively.

\textbf{CNN}. As shown in Figure \ref{fig:layout:kernel}, the structure of CNN adopted in the experiments is the same as that of QCNN, with the quantum convolutional kernels substituted by classical convolutional kernels and other components unchanged.

\begin{figure}[htp]
    \centering
    \includegraphics[scale=0.7]{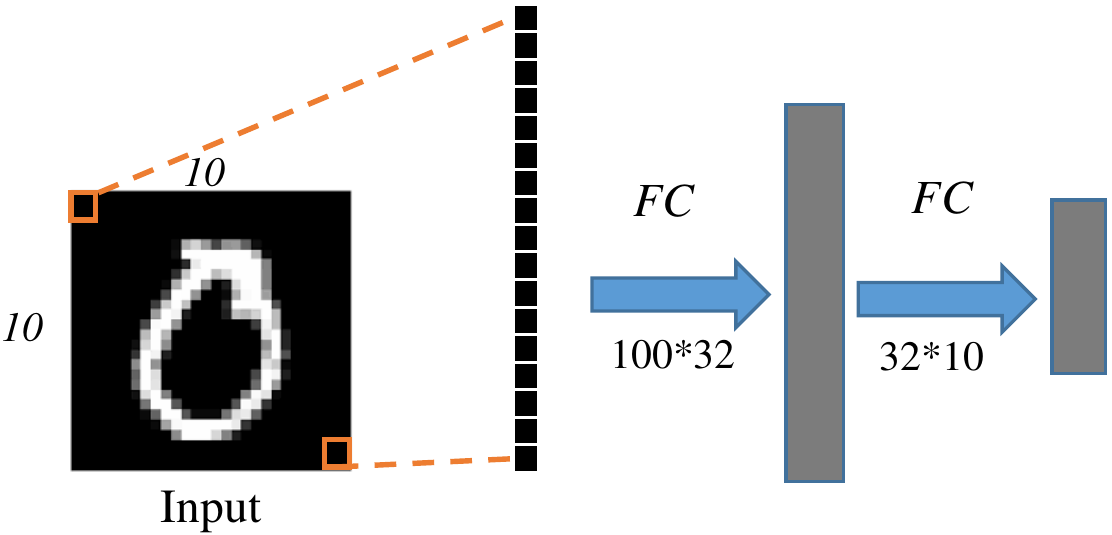}
    \caption{\textbf{Layout of MLP for MNIST}. We first convert the image to an 1-D vector, followed by two FC (fully-connected) layers. By only replacing the first convolutional layer of CNN and QCNN with a FC layer, we can figure out the role of quantum convolutional kernel in the training.}
    \label{fig:layout:mlp}
\end{figure}

\textbf{MLP}. MLP is constructed by sequentially connecting multiple fully-connected layers one by one. When processing quantum synthetic data and the Wine data, we adopt a three-layer MLP, with the dimension of hidden layer depending on the limitation of total number of trainable parameters. For MNIST dataset, the original 2-D image is firstly flattened into an 1-D vector, which then input into a three-layer MLP with 32 hidden nodes, as described in Figure \ref{fig:layout:mlp}. Therefore, the MLP designed for MNIST replaces the convolution layer by the fully-connected layer. In this way, we can observe how quantum convolution affects the training process.

\subsection{The deployed hardware and hyper-parameter settings}
The QNNs referred in this paper are implemented based on Pennylane \cite{bergholm2018pennylane} and PyTorch \cite{NEURIPS2019_9015}. Due to the limited accessibility of physical quantum computers, all experiments are simulated on classical computers with Intel(R) Xeon(R) Gold 6267C CPU @ 2.60GHz and 128 GB memory. To simulate the quantum noise, we adopt the noise model from ibmq\_16\_melbourne, which is one of the IBM Quantum Canary Processors \cite{IBMQ}. 

For fair comparison, the number of trainable parameters of different models for the same task is kept close and the structure-independent hyper-parameters during training are not fine tuned specifically for each model but always keep the same setting for one task. Meantime, the experiment with a specific configuration is repeated 10 times and calculate the average as final experiment result. The rigorous experiment setup is listed in Table \ref{tab:hyperparam}.

\begin{table}[htp]
\centering
\caption{Hyper-parameter setting in the experiments}
\label{tab:hyperparam}
\begin{tabular}{llll}
\toprule
Model  &   Learning rate &   Batch size &  Optimizer\\
\midrule
QNNN   & 0.01 & 4 & SGD \\
QENN  & 0.01 & 4 & SGD \\
QCNN  & 0.001  & 5 & Adam \\
CNN & 0.001 & 5 & Adam \\
\bottomrule
\end{tabular}
\end{table}

\section{The performance of QNNs in the NISQ scenario}\label{append:result}
\begin{figure*}[htp]
    \centering
    \includegraphics[scale=0.6]{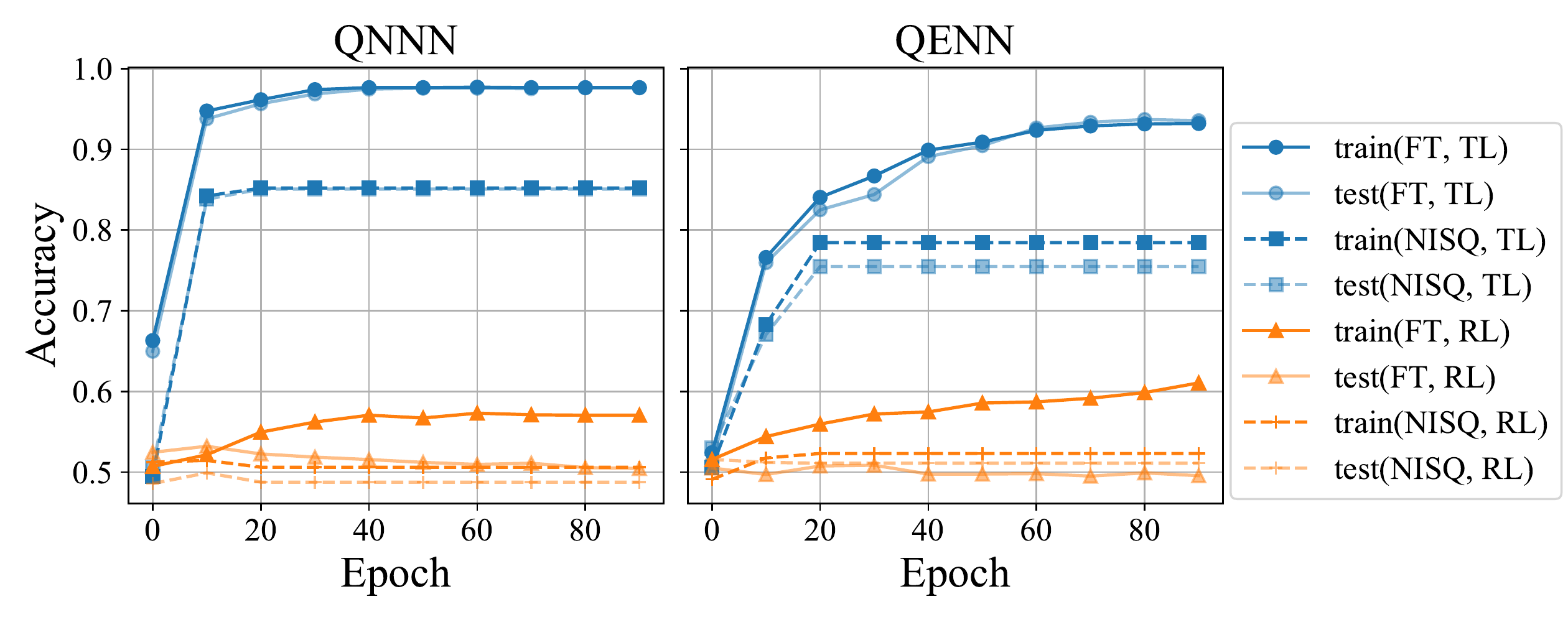}
    \caption{\small{\textbf{Quantum noise reduces quantum model capacity.} \textsf{FT} denotes the fault-tolerant devices, \textsf{TL} is the abbreviation of "true label", and \textsf{RL} is the abbreviation of "random label". When running the same quantum circuit on noisy devices,there is a serious performance decline in terms of the ability of learning the original data and fitting random label.}}
    \label{fig:performance:NISQ}
\end{figure*}

In this section, we explore the performance of QNNs under the NISQ scenario. The achieved results are depicted in Figure \ref{fig:performance:NISQ}. In particular, the solid and dashed lines describe the training process on noiseless and NISQ devices respectively. An immediate observation is that quantum system noise largely weaken the power of quantum models, leading to a severe accuracy drop (from about $94\%$ to $80\%$) on quantum synthetic data. Furthermore, the noisy QNNs seems to completely lose the ability of matching the random data.

\section{Trainability of QNNs on other datasets}\label{append:trainability}
\begin{figure*}[htp]
    \centering
    \includegraphics[scale=0.6]{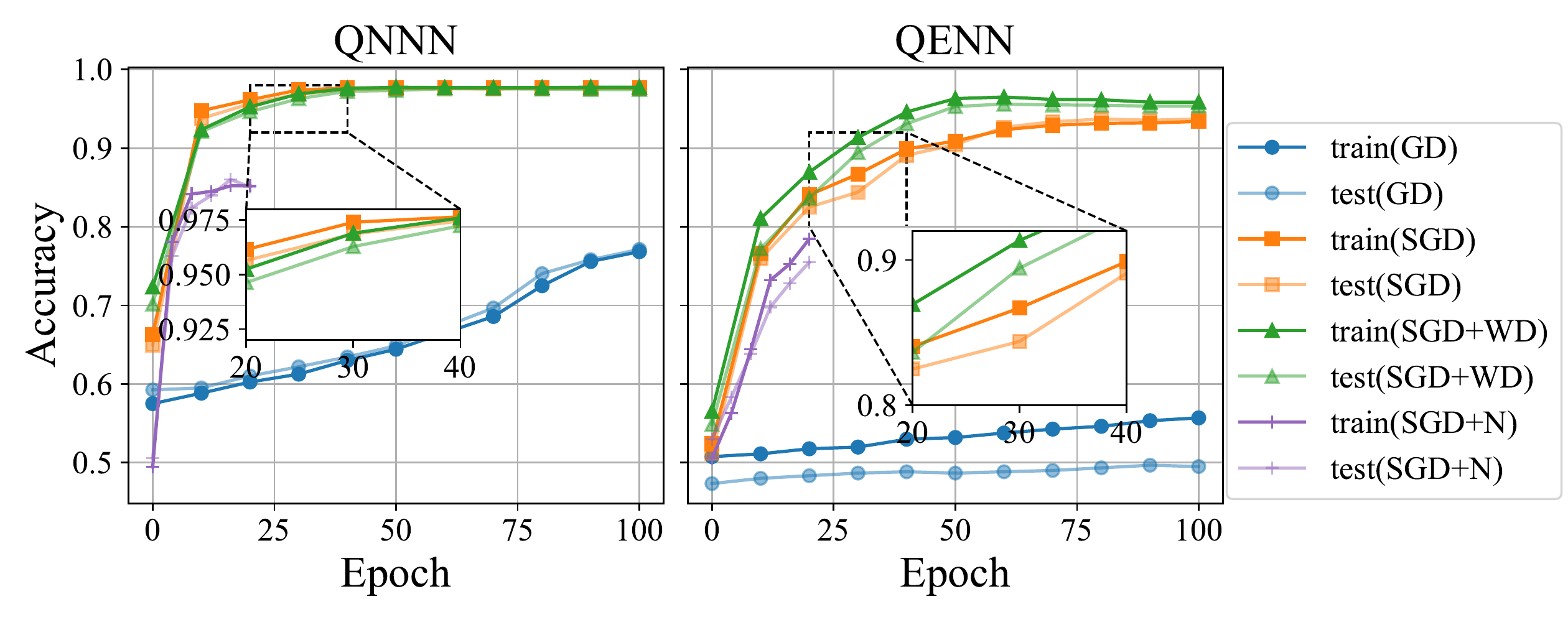}
    \caption{\small{\textbf{Effects of regularization techniques on the training of QNNs on quantum synthetic data.} \textsf{GD} is gradient decent, \textsf{SGD} is stochastic gradient descent, \textsf{WD} is weight decay and \textsf{N} means running the experiments on NISQ chips. SGD plays a significant role in accelerating convergence and achieving higher accuracy, while early stopping hinders the optimization process instead of boosting performance.}}
    \label{fig:general}
\end{figure*}

Here we proceed a series of experiments investigate the trainability of QNNs with respect to the two regularization techniques, i.e., stochastic gradient descent (SGD) and weight decay, on the quantum synthetic dataset. The collected results are shown  in Figure \ref{fig:general}. 

\textbf{SGD.} The batch SGD optimizer effectively boosts the convergence rate of QNNs without obvious over-fitting. Specifically, in the first 20 epochs, the train and test accuracy of QNNN (QENN) with SGD rapidly rise to nearly $95\%$ ($80\%$), while GD optimizer only helps QNNN (QENN) achieve $3\%$ ($2\%$) accuracy growth with respect to the initial models. After finishing the whole training process after 100 epochs, there is still a $20\%$ ($30\%$) accuracy gap between QNNN (QENN) optimized by the vanilla GD and SGD optimizers.

\begin{figure*}[htp]
    \centering
    \subfigure[Training process]{
    \includegraphics[scale=0.38]{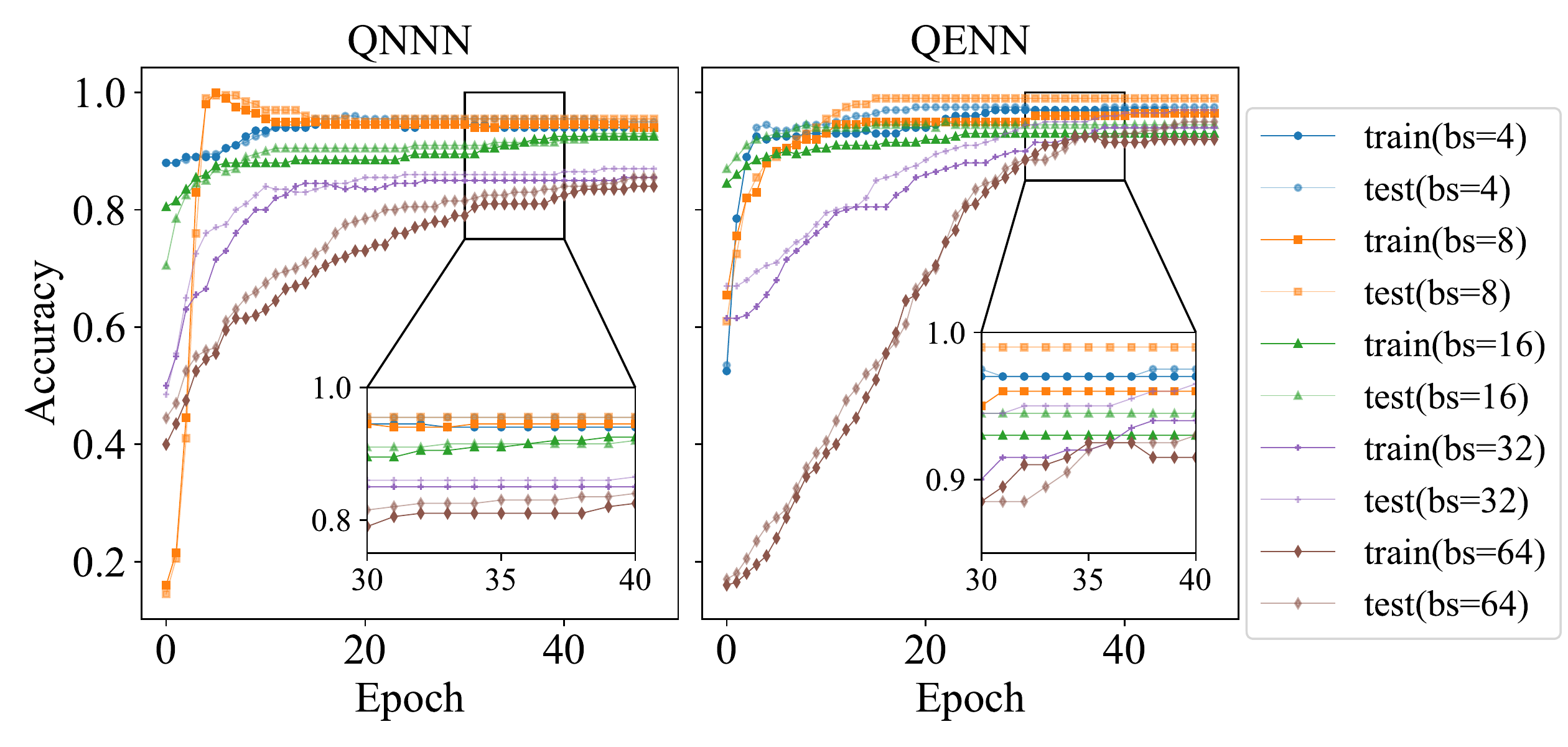}}
    \subfigure[Accuracy vs batch size]{
    \includegraphics[scale=0.38]{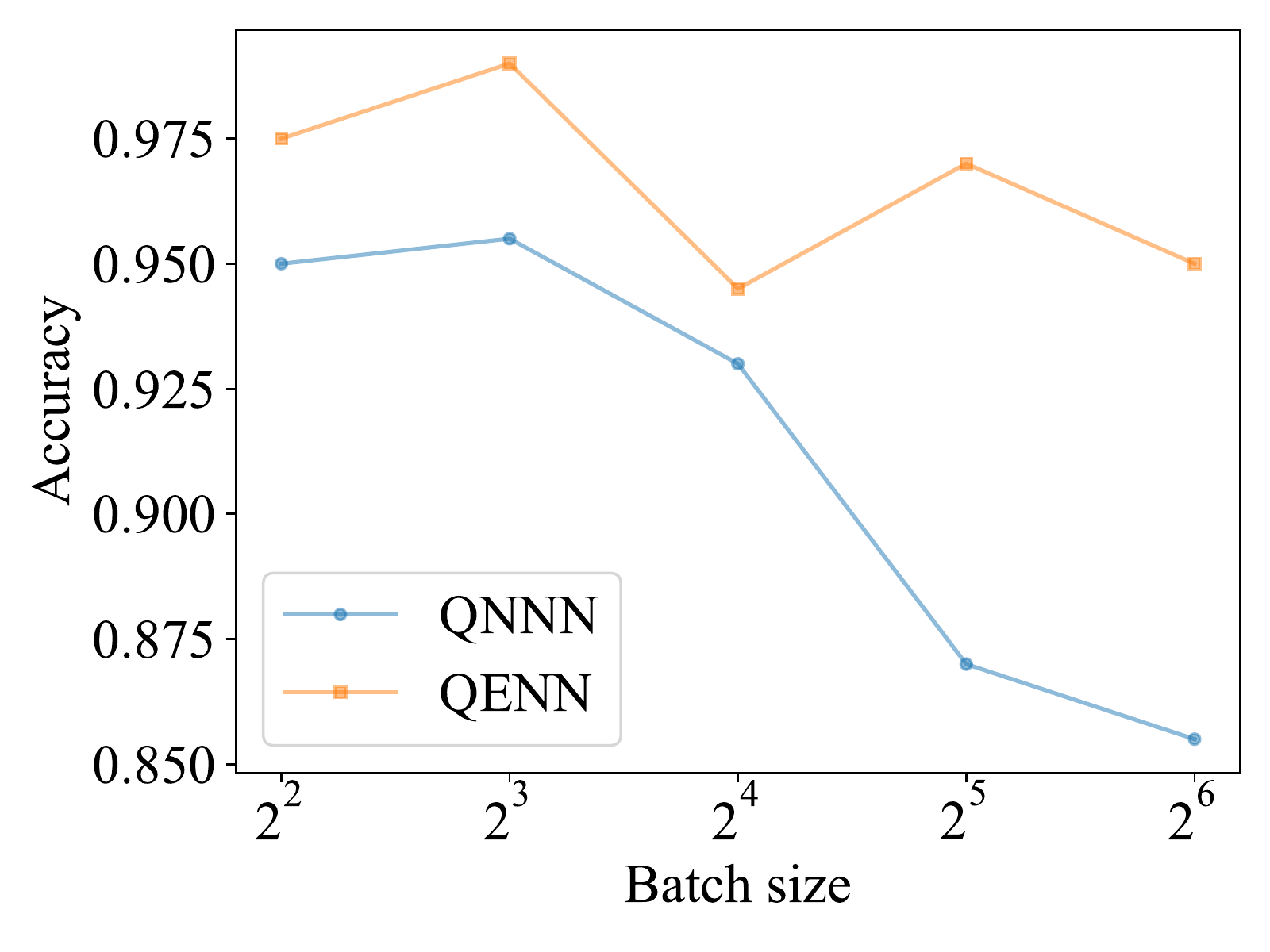}}
    \caption{\small{\textbf{Batch size of SGD significantly influences the training of quantum models.} (a) shows the convergence of quantum models with different setting of batch size. \textsf{bs} is abbreviation of batch size. (b) shows the the fluctuation of test accuracy with respective to batch size.}}
    \label{fig:batchsize}
\end{figure*}

In view of the substantial positive effects brought by SGD on the training of QNNs, we further exploit how the batch size effects the trainability of QNNs. As depicted in Figure \ref{fig:batchsize}, it appears an nonmonotonic decline of the convergence rate and accuracy with respect to the increment of batch size. And there exists an optimal batch size, which allows the best convergence rate and the highest accuracy (the best batch size is around 8 in our setting).

\textbf{NISQ.} Although SGD optimizer  facilitates the trainability of QNNs, the system noise in NISQ devices weakens the positive effect brought by SGD. As indicated by the purple line in Figure \ref{fig:general}, it appears almost $10\%$ ($5\%$) accuracy decay for QNNN (QENN) in the first 20 training epochs in the NISQ scenario. Meantime, the existence of noise extremely amplifies the runtime of QNNs by classical simulations (Please refer to Appendix \ref{append:runtime} for more details).

\textbf{Weight decay. }Weight decay plays a slightly different role in the training of QNNN and QENN. For QNNN, there is no significant effect after applying the weight decay regularizer to the SGD optimizer. For QENN which is more powerful than QNNN, the strategy of weight decay leads to faster convergence and finally achieve a higher accuracy. Note that both QNNN and QENN do not encounter apparent over-fitting issue on quantum synthetic data, which leaves little space for weight decay to show its potential in alleviating over-fitting.

\section{Runtime analysis}\label{append:runtime} 
\begin{figure*}[htp]
    \centering
    \subfigure[]{
    \includegraphics[scale=0.36]{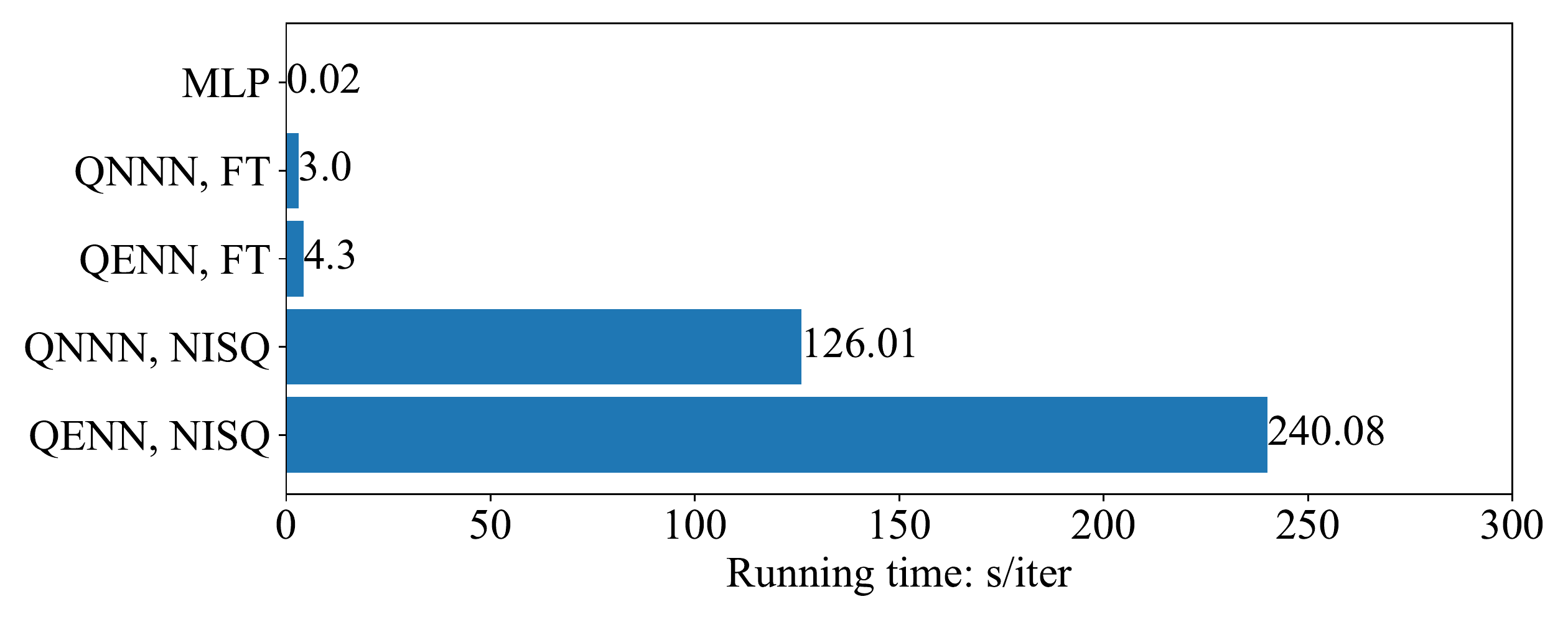}}
    \subfigure[]{
    \includegraphics[scale=0.36]{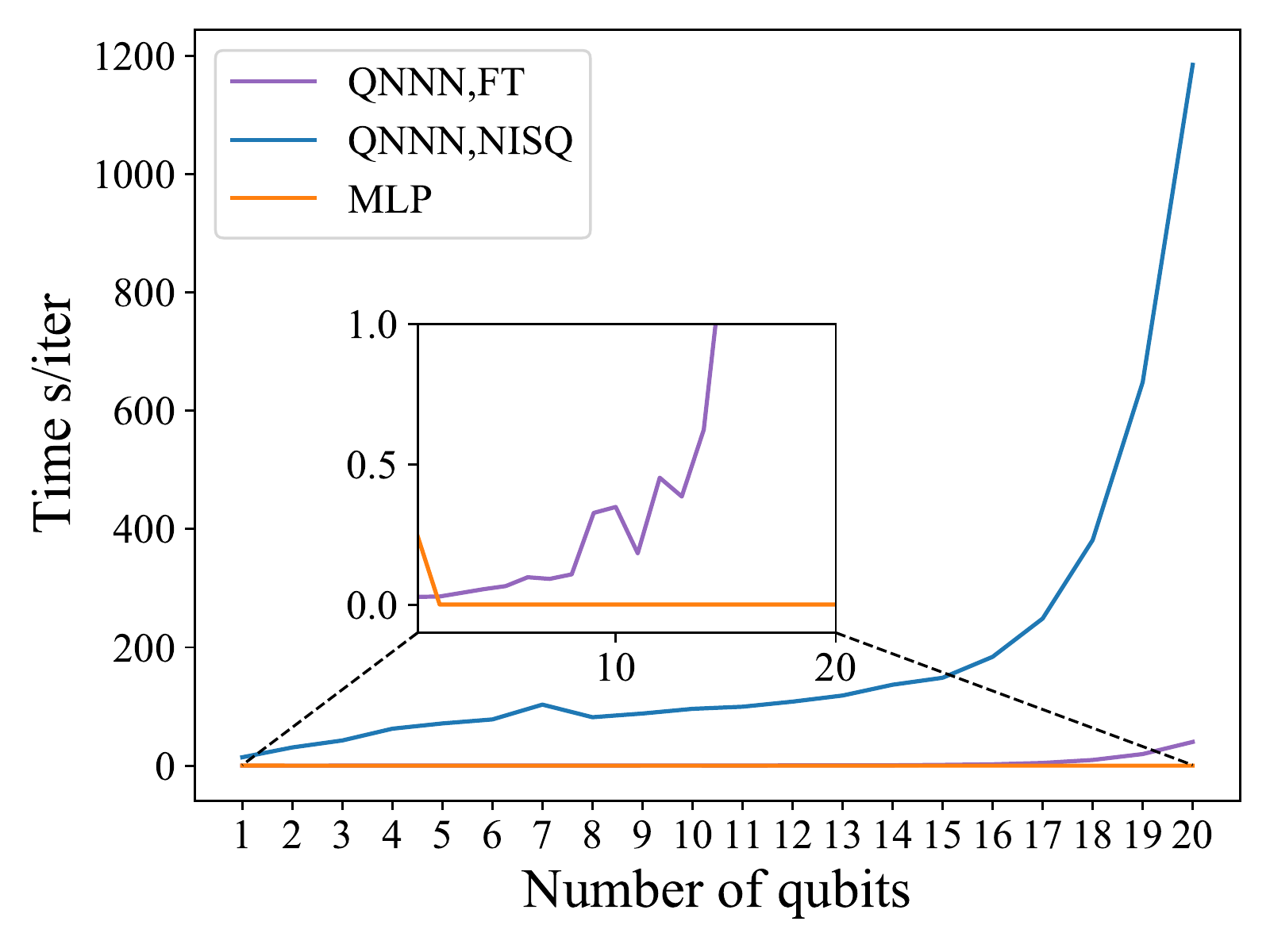}}
    \caption{\small{\textbf{Runtime of QNNs and DNNs simulated by classical computers.} \textsf{FT} is the abbreviation of "Fault-tolerant". (a) Comparison of running time of every iteration for each setting. Currently, the time cost of simulating noisy quantum circuits on classical computers is unacceptable, which is hundreds of times higher than that of the counterpart of classical version. (b) With the increasing number of required qubits for feature embedding, the time of training noisy QNNN grows exponentially while MLP sees a slow and negligible growth of running time.}}
    \label{fig:runtime}
\end{figure*}

We compare the runtime of QNNs and DNNs when processing the Wine dataset with the fixed number of qubits. As shown in Figure \ref{fig:runtime} (a), simulating QNNs under the NISQ setting suffers from the lowest efficiency, which averagely spends $126$s to execute one iteration. For the fault-tolerant setting, the running time of every optimization step steeply declines to $3$s. However, there still exists a huge gap of time cost between simulating QNNs and DNNs on classical computers. Another noticeable phenomenon is that noise cause a greater negative impact on the running time of larger-scale QNNs. After applying noise in the training, QENN spends more $114$s than QNNN on every iteration, while the increment in time with fault-tolerant setting is $1$s. 

In light of the exponential growth of runtime with the increased scale of QNNs, we next indicate how the number of qubits effects their simulation time. As shown in Figure \ref{fig:runtime} (b),  the runtime of simulating QNNN is sensitive to the qubit scale, appearing an exponential growth with the increment of qubits, especially for QNNN in NISQ environment. On the contrary, MLP experiences a negligible growth of time on every training iteration when the dimensions of input feature vector and hidden layer steadily increase.

\end{document}